\documentclass[11pt]{article}
\usepackage{fullpage}
\usepackage[table,xcdraw]{xcolor}
\usepackage[margin=1in]{geometry}
\usepackage{xfrac}
\usepackage{flushend}
\usepackage{amsmath}
\usepackage{amsthm}
\usepackage{amssymb}
\usepackage{caption}
\usepackage{bm}
\usepackage{multicol}
\usepackage[style=alphabetic,minalphanames=3,natbib=true,maxbibnames=99]{biblatex}
\usepackage{enumitem}
\usepackage{wrapfig}
\usepackage{worldflags}
\usepackage{multirow}
\usepackage{mathtools}
\AtBeginDocument{%
  \providecommand\BibTeX{{%
    \normalfont B\kern-0.5em{\scshape i\kern-0.25em b}\kern-0.8em\TeX}}}

\usepackage{algorithm}
\usepackage{xspace}
\usepackage[noend]{algpseudocode}
\usepackage{nicefrac}
\usepackage{appendix}
\usepackage{lipsum}
\usepackage{geometry}
\usepackage[colorlinks]{hyperref}

\hypersetup{
    linkcolor = {violet},
    citecolor = {teal},
    urlcolor = {teal}
}

\usepackage[]{algpseudocode}

\RequirePackage[T1]{fontenc} 
\RequirePackage[tt=false, type1=true]{libertine} 
\RequirePackage[varqu]{zi4} 
\RequirePackage[libertine]{newtxmath}

\usepackage{wrapfig}

\theoremstyle{plain}
\newtheorem{theorem}{Theorem}[section]
\newtheorem{thm}[theorem]{Theorem}
\newtheorem{lem}[theorem]{Lemma}

\newtheorem{dfn}[theorem]{Definition}

\newtheorem{ass}[theorem]{Assumption}

\DeclareMathOperator*{\argmin}{arg\,min}

\newcommand{\DAIO}{\ensuremath{\mathtt{DAIO}}\xspace}
\newcommand{\FOIO}{\ensuremath{\mathtt{FOIO}}\xspace}
\newcommand{\ZOIO}{\ensuremath{\mathtt{ZOIO}}\xspace}

\newcommand{\mbf}[1]{{\mathbf{#1}}}

\newcommand{\sref}[2]{\hyperref[#2]{#1 \ref{#2}}}

\definecolor{ao(english)}{rgb}{0.0, 0.5, 0.0}

\usepackage[textsize=tiny]{todonotes}

\usepackage{etoolbox}
\makeatletter
\patchcmd{\hyper@makecurrent}{%
    \ifx\Hy@param\Hy@chapterstring
        \let\Hy@param\Hy@chapapp
    \fi
}{%
    \iftoggle{inappendix}{%
        \@checkappendixparam{chapter}%
        \@checkappendixparam{section}%
        \@checkappendixparam{subsection}%
        \@checkappendixparam{subsubsection}%
        \@checkappendixparam{paragraph}%
        \@checkappendixparam{subparagraph}%
    }{}%
}{}{\errmessage{failed to patch}}

\newcommand*{\@checkappendixparam}[1]{%
    \def\@checkappendixparamtmp{#1}%
    \ifx\Hy@param\@checkappendixparamtmp
        \let\Hy@param\Hy@appendixstring
    \fi
}
\makeatletter

\newtoggle{inappendix}
\togglefalse{inappendix}

\apptocmd{\appendix}{\toggletrue{inappendix}}{}{\errmessage{failed to patch}}

\newlength{\whilewidth}
\algdef{SE}[parWHILE]{parWhile}{EndparWhile}[1]
{\parbox[t]{\dimexpr\linewidth-\algmargin}{%
		\hangindent\whilewidth\strut\algorithmicwhile\ #1\ \algorithmicdo\strut}}{\algorithmicend\ \algorithmicwhile}%
\algnewcommand{\parState}[1]{\State%
	\parbox[t]{\dimexpr\linewidth-\algmargin}{\strut #1\strut}}
\makeatother

\title{Optimizing Individualized Incentives from Grid Measurements and Limited Knowledge of Agent Behavior}

\author{
  Adam Lechowicz \thanks{National Renewable Energy Laboratory \& University of Massachusetts Amherst.  Email: \url{alechowicz@cs.umass.edu}}
  \and
  Joshua Comden \thanks{National Renewable Energy Laboratory.  Email: \url{joshua.comden@nrel.gov}}
  \and
  Andrey Bernstein \thanks{National Renewable Energy Laboratory.  Email: \url{andrey.bernstein@nrel.gov}}
}

\begin{document}

\maketitle

\begin{abstract}
As electrical generation becomes more distributed and volatile, and loads become more uncertain, controllability of distributed energy resources (DERs), regardless of their ownership status, will be necessary for grid reliability.
Grid operators lack direct control over end-users' grid interactions, such as energy usage, but \textit{incentives} can influence behavior -- for example, an end-user that receives a grid-driven incentive may adjust their consumption or expose relevant control variables in response.

A key challenge in studying such incentives is the lack of data about human behavior, which usually motivates strong assumptions, such as distributional assumptions on compliance or rational utility-maximization.  
In this paper, we propose a general incentive mechanism in the form of a constrained optimization problem -- our approach is distinguished from prior work by modeling human behavior (e.g., reactions to an incentive) as an arbitrary unknown function.  
We propose feedback-based optimization algorithms to solve this problem that each leverage different amounts of information and/or measurements.
We show that each converges to an asymptotically stable incentive with (near)-optimality guarantees given mild assumptions on the problem.
Finally, we evaluate our proposed techniques in voltage regulation simulations on standard test beds.  We test a variety of settings, including those that break assumptions required for theoretical convergence (e.g., convexity, smoothness) to capture realistic settings.  In this evaluation, our proposed algorithms are able to find near-optimal incentives even when the reaction to an incentive is modeled by a theoretically difficult (yet realistic) function.
\end{abstract}
\clearpage

\section{Introduction}
\label{sec:intro}

With both the rapid adoption of volatile renewable energy resources replacing fossil fuel power plants throughout the power grid and ever-changing load patterns due to electrification of broad sectors, maintaining reliable power service is becoming increasingly difficult for power system operators~\cite{NERC:24}; increasing controllability on both the supply and demand sides of the grid will be necessary to deal with any unexpected mismatches between the two.
While the sources of uncertainty from renewable energy resources are well understood, the complexity and diversity among loads make understanding how to control them much more burdensome, especially amongst those loads that are closely controlled by humans or, more generally, end-users.

In this work, we consider how a system operator (e.g., of an electrical distribution grid) can gain controllability over assets not owned by them (whether direct or indirect) through the use of \textit{incentives}.  As the adoption of assets such as distributed energy resources (DERs) on the grid edge grows, the importance of controllability over such assets for distribution grid reliability grows as well, especially with demand-side assets~\cite{Deloitte:24,Brattle:19} and geographically targeting the incentives~\cite{Carvallo:23}. As an example of an incentive-based program, the municipal electrical utility of San Antonio, Texas will give customers up to \$50/year to allow the utility to directly adjust their building's temperature by 2-3$^{\circ}$F during peak usage times that amount to 65 MW of distributed load reduction~\cite{Carnett:23}.  %

The goal of these control efforts is to provide asset owners with an incentive (e.g., via price signals or direct incentive payments) in exchange for adjusting their asset towards an operator's desired set point.  From the system operator's perspective, their objective is to minimize the total incentive that they are required to provide, subject to the desired safety constraints.
However, the nature of this problem brings significant challenges due to a lack of knowledge about the asset owners' sensitivity to incentives (i.e., how they respond when given a certain incentive), which is heterogeneous and non-stationary.  The system operator may have estimates of this sensitivity (e.g., based on survey data and load models~\cite{Asadinejad:18}, or historical data~\cite{Bian:22}), but depending on the characteristics of a specific setting or user group, this may not be the case.

Existing studies that consider this topic have approached the above challenge in a few different ways, including modeling agents as rational utility-maximizers with respect to energy prices~\cite{Cavraro:24}, assuming that agents comply with grid signals according to a probability distribution~\cite{Cui:23}, applying a contract that is linearly dependent on grid conditions~\cite{Comden:17}, having an aggregator that can accept offers from prosumers on their flexibility~\cite{Lilliu:23}, dynamic pricing of demand response under the assumption of a concave utility function of the prosumers~\cite{Zhao:23}, or price-based flexibility contracts with competitive guarantees~\cite{Xu:22}.

In what follows, we consider a setting with a general individualized incentive mechanism that places few assumptions on the response of asset owners; such a setting is necessary to simultaneously model many different potential incentivization schemes (i.e., in the mechanism) and diverse environments where out-of-distribution effects and hard-to-predict human behavior would challenge assumptions made in prior work (i.e., in the incentive response).
The new challenges brought on by this problem setting require different algorithmic techniques to obtain incentives that satisfy system constraints while optimizing the system operator's objectives.  Thus, in this work, we consider the following question:

\begin{center} 
\textit{Is it possible to efficiently optimize individual incentives for demand-side control under system stability constraints where stakeholder responses are arbitrary and difficult to predict}?
\end{center}

\smallskip
\noindent\textbf{Contributions. \ } 
In answering the question posed above, we make two primary contributions.  First, we introduce a general model of incentive response that can capture many possible incentivization schemes and arbitrary behavior (e.g., from end-users of a system).  This model, which generalizes settings considered in prior work, may be of independent interest beyond this work and the motivating setting of power grids.  
Answering the question in the affirmative, we propose iterative algorithms (see \autoref{sec:opt}) that optimize the incentive amounts given by the system operator in a feedback-based manner.  These algorithms are simple to implement and are designed for different knowledge scenarios that depend on how much incentive response behavior information is known by the system operator. We show proportionately strong theoretical results for each technique, and we give a case study, where we implement the proposed techniques to optimize load shedding for voltage control on a simulated distribution grid.  We experimentally show that the proposed algorithms are empirically useful even in ``realistic'' settings where the problem lacks many of the features required for theoretical bounds (e.g., differentiability, convexity).

\smallskip
\noindent\textbf{Notation. \ }
Throughout the paper, lowercase bold letters denote vectors, and uppercase bold letters denote matrices.  The identity matrix, all-ones vector, and all-zeros vector are denoted by $\mathbf{I}$, $\mathbf{1}$, and $\mathbf{0}$ (respectively), where the dimension is inferred based on context.  The set of real numbers and non-negative real numbers in $n$-dimensions are denoted by $\mathbb{R}^n$ and $\mathbb{R}^n_{+}$, respectively, and $\left[ \cdot \right]_{\mathbb{R}^n_{+}}$ denotes \textit{projection} to the positive orthant.  The operators $\succcurlyeq$ and $\preccurlyeq$ are used to denote element-wise inequalities ($\geq$ and $\leq$, respectively) that compare the elements of two same-length vectors.  Likewise, we use the Hadamard notation of $\odot$, $\oslash$, and $\square^{\circ \square}$ to denote element-wise multiplication, division, and exponentiation of vectors, respectively.

\section{Problem Formulation and Preliminaries}
\label{sec:prob}
In this section, we begin by formulating the \textit{system control with incentives} problem as a general constrained optimization problem.  We then instantiate this problem towards the motivating application of distribution grid voltage control, and give some mild structural assumptions that will inform our algorithm design in \autoref{sec:opt}.

\subsection{System Control with Incentives}

Suppose that a system operator (SO) desires to set an operating point $\mbf{u}^\star$ of the resources that it does not directly control.
To do this, it sends incentives $\mbf{i}$ along with the operating point $\mbf{u}^\star$ to the resource owners which comes at a cost $c(\mbf{i})$ incurred by the SO.
It will be the discretion of the resource owners to how close to $\mbf{u}^\star$ the resource operating points are set which depends on the amount of incentives $\mbf{i}$ given to them; in general, the larger incentives given, the closer the resources are set to operating at $\mbf{u}^\star$.

The system's measured output is denoted by the function $h(\mbf{u})$ for operating point $\mbf{u}$, where $h(\mbf{u}) \preccurlyeq \mbf{0}$ denotes that the system is operating safely.  The SO has knowledge of the function $h(\cdot)$, but may have to measure the current operating point $\mbf{u}$.

Given the SO's desired operating point for the system $\mbf{u}^\star$ and a certain incentive $\mbf{i}$, the environment (e.g., resource owners actions) responds by adjusting the system's operating point according to an \textit{unknown} function $g_{\mbf{u}^\star}(\mbf{i})$ that represents the implemented operating point.  Although this function is unknown, the SO can measure (sample) outcomes, i.e., can see $g_{\mbf{u}^\star}(\mbf{i}) \rightarrow \mbf{u}$ for a specific $\mbf{i}$ input.

Using $h(\cdot)$ as a \textit{system safety constraint}, in \eqref{eq:sci} we present the optimization problem that finds the best (i.e., lowest cost) incentive that maintains system safety.
\begin{align}
&\begin{aligned}
\min_{\mbf{i} \in \mathbb{R}^n} & \ \ c(\mbf{i}),  \quad (\textit{cost of incentive})\\
\text{s.t., } & \ \ h(g_{\mbf{u}^\star}(\mbf{i})) \preccurlyeq \mbf{0}. \quad (\textit{system constraints}) 
\end{aligned} \label{eq:sci}
\end{align}

\noindent It is assumed that the SO specifies a desired operating point $\mbf{u}^\star$ that satisfies the system constraints, i.e., $h(\mbf{u}^\star) \preccurlyeq \mbf{0}$ -- otherwise, the problem will be infeasible.

\subsection{Distribution Grid Voltage Control} \label{sec:distgridVC}

In this section, we instantiate the optimization problem \eqref{eq:sci} towards the motivating application of voltage control in distribution grids.

An electric distribution grid with $n+1$ buses can be modeled by an undirected graph $G = (\mathcal{N}, \mathcal{E})$, where nodes $\mathcal{N} = \{ 0, 1, ..., n \}$ are associated with the electrical buses and edges $\mathcal{E}$ represent electric lines.
The substation is labeled as $0$ and is modeled as an ideal voltage generator (i.e., a slack bus), imposing a nominal voltage of 1 p.u.  All other buses (i.e., the PQ buses) are assumed to be \textit{prosumers}, meaning that they are both a producer and a consumer of energy.  The $j^{\text{th}}$ prosumer generates \textit{active power} $r_j \in \mathbb{R}_+$ due to e.g., ``behind-the-meter'' energy resources such as residential PV.  Prosumer $n$ also has active and reactive power demands $d_j \in \mathbb{R}_+$ and $q_j \in \mathbb{R}$, respectively.  The \textit{net} active power injection (or absorption) of prosumer $j$ is given by $p_j = r_j - d_j$.
Active and reactive powers take positive values ($p_j, q_j \geq 0$) when they are injected into the grid, and $j$ behaves like a generator.   Otherwise, when powers take negative values ($p_j, q_j \leq 0$), they are absorbed from the grid, and $j$ behaves like a load.  We will henceforth let $\mbf{p} \in \mathbb{R}^n, \mbf{q} \in \mathbb{R}^n$ collect all of the active and reactive power values at PQ buses.

We denote by $v_j \in \mathbb{R}$ the \textit{voltage magnitude} at bus $j \in [n]$, and let the vector $\mbf{v} \in \mathbb{R}^n$ collect the voltages of buses $1, ..., n$.  In general, voltage magnitudes
can be solved for with the non-linear power flow equations of the system -- however, the canonical LinDistFlow model~\cite{Baran:89} gives that a first-order Taylor expansion of the power flow equations that yields the following approximation around a linearization point of $\mbf{p}^\star, \mbf{q}^\star, \mbf{v}^\star$:
\begin{equation}
    \mbf{v} = \mbf{Rp} + \mbf{Xq} + \widetilde{\mbf{v}},
\end{equation}
where $\mbf{R} \in \mathbb{R}^{n \times n}_+$ and $\mbf{X} \in \mathbb{R}_+^{n \times n}$ are symmetric and positive definite matrices~\cite{Cavraro:22}, and $\widetilde{\mbf{v}} \in \mathbb{R}^n_+$.  They are defined as follows:
\begin{align*}
    \mbf{R} &= 2 \text{Re} \left( \text{diag}(e) \overline{\mbf{Z}} \text{diag}(e)^{-1} \right),\\
    \mbf{X} &= - 2 \text{Im} \left( \text{diag}(e) \overline{\mbf{Z}} \text{diag}(e)^{-1} \right),\\
    \widetilde{\mbf{v}} &= \mbf{v}^\star - \mbf{Rp}^\star - \mbf{Xq}^\star.
\end{align*}
$\overline{\mbf{Z}}$ is the complex conjugate of the inverse of the reduced admittance matrix $\mbf{Y}_{LL}$ (the admittance matrix omitting the slack bus), and $e$ is the open circuit voltage, defined as $e \coloneqq -\mbf{Y}_{LL}^{-1} \mbf{Y}_{LS} \mbf{V}_{S}$, where $\mbf{V}_{S}$ is the fixed voltage phasor at the slack bus.  We henceforth let $\mbf{u}^\star$ (i.e., the SO's setpoint in \eqref{eq:sci}) and $\mbf{u}$ (i.e., the operating point defined by $g_{\mbf{u}^\star}(\mbf{i})$) implicitly indicate a corresponding tuple of PQ values, e.g., $\mbf{u}^\star \coloneqq (\mbf{p}^\star, \mbf{q}^\star)$ and $\mbf{u} \coloneqq (\mbf{p}, \mbf{q})$.
Suppose that the system operator defines bounds on the load voltage magnitudes, given by $\overline{\mbf{v}}$ and $\underline{\mbf{v}}$.
The system safety constraint is then given by
\[
\underline{\mbf{v}} \preccurlyeq \mbf{v} \preccurlyeq \overline{\mbf{v}}.
\]

In terms of the LinDistFlow model and the form of Problem \eqref{eq:sci}, we define $h(\mbf{u})$ for voltage control as follows.  Recall that $\mbf{u}$ (the current operating point) implicitly indicates the tuple $(\mbf{p}, \mbf{q})$:
\begin{equation}
    h(\mbf{u}) \coloneqq \max\{ \underline{\mbf{v}} - \left(\mbf{Rp} + \mbf{Xq} + \widetilde{\mbf{v}}\right), \left( \mbf{Rp} + \mbf{Xq} + \widetilde{\mbf{v}} \right) - \overline{\mbf{v}}  \}, \label{eq:h-func}
\end{equation}
where note that $h(\mbf{u}) \preccurlyeq \mbf{0}$ implies that the current operating voltages satisfy the bounds defined by $[\underline{\mbf{v}}, \overline{\mbf{v}}]$.

In what follows, it will be useful to consider a simplified form of $h(\cdot)$ that arises when, e.g., we only expect the voltage lower bound to be a concern (e.g., all PQ buses are consumers, not prosumers), and reactive power is not responsive to incentives (e.g., due to inverter-based DERs).  Letting $\mbf{u}$ indicate $\mbf{p}$, we have the following:
\begin{equation}
h(\mbf{u}) \coloneqq \underline{\mbf{v}} - \mbf{Ru} - \mbf{Xq} - \widetilde{\mbf{v}} \label{eq:simple-h}   
\end{equation}

\smallskip
\noindent\textbf{Assumptions. \ }
In the rest of the paper, we make the following mild assumptions on Problem \eqref{eq:sci} motivated by the empirical structure of the motivating application.
First, we let $c(\mbf{i}) \coloneqq \Vert \mbf{i} \Vert_1$ -- i.e., the cost of the incentive is given by a simple $\ell_1$ norm.\footnote{ We comment that $c(\mbf{i})$ could theoretically be a more complex function of $\mbf{i}$ and/or $g_{\mbf{u}^\star}(\mbf{i})$, e.g., if the SO's cost depends on the end-user's response.  The feedback-based algorithms we present in \autoref{sec:opt} mostly rely on the sensitivities given by $\partial_{\mbf{i}} c(\mbf{i})$, which require only minor modifications from an algorithmic standpoint.  We defer a deeper exploration of more complicated incentive cost functions to future work. }

\begin{ass}[Monotonicity] \label{asm:monotone}
As the elements of incentive $\mbf{i}$ grow away from $\mbf{0}$, the operating point $\mbf{u} = g_{\mbf{u}^\star}(\mbf{i})$ monotonically approaches the SO's desired setpoint $\mbf{u}^\star$.  

Formally, for any two non-negative incentives $\mbf{i}^{(1)}$ and $\mbf{i}^{(2)}$ where $\mbf{i}^{(1)} \succcurlyeq \mbf{i}^{(2)}$, we assume that $\Vert g_{\mbf{u}^\star}(\mbf{i}^{(1)}) - \mbf{u}^\star \Vert \leq \Vert g_{\mbf{u}^\star}(\mbf{i}^{(2)}) - \mbf{u}^\star \Vert$.
\end{ass}
\noindent In a real application such as voltage control, \sref{Assumption}{asm:monotone} corresponds to the idea that increasing the SO's incentive does not drive the system \textit{away} from the safe point specified by the SO; in the worst-case, such an increase in the incentive can have no effect.
\begin{ass}[Threshold vector $\mbf{t}$] \label{asm:threshold}
Without loss of generality, we assume that there exists a vector of incentive \textit{thresholds} $\mbf{t} \succcurlyeq \mbf{0}$ at which the environment sets the operating point equal to the setpoint $\mbf{u}^\star$.  Formally, we have:
\[
g_{\mbf{u}^\star}(\mbf{t}) = \mbf{u}^\star.
\]
Each element of $\mbf{t}$ is allowed to be arbitrarily large but finite.
\end{ass}

\sref{Assumption}{asm:threshold} encodes the empirically realistic idea that there exists some (potentially very large) incentive for which the environment will comply with the system operator's desired setpoint.  Observe that if this assumption does not hold, it is equivalent to say that there exists some user(s) who is unwilling to meet the SO's setpoint for \textit{any} incentive, which might cause Problem \eqref{eq:sci} to be infeasible.  In what follows, we leverage this assumption to define a linear approximation of any arbitrary $g$ function that we will use in our theoretical analysis and experiments.

\begin{dfn}[Linear approximation $\overline{g}_{\mbf{u}^\star}( \mbf{i})$] \label{dfn:linear-g}
Given a true function $g_{\mbf{u}^\star}(\mbf{i})$, we henceforth use vector $\bm{\delta}$ to represent the element-wise deviance from $\mbf{u}^\star$ when the incentive is zero, i.e.,
\[
\bm{\delta} \coloneqq \overline{g}_{\mbf{u}^\star}( \mbf{0}) - \mbf{u}.
\]
Recall \sref{Assumption}{asm:threshold}.  Given a threshold vector $\mbf{t}$ and the value of $\bm{\delta}$, we can construct a \textit{linear approximation} of the true function $g_{\mbf{u}^\star}(\mbf{i})$; we let $\overline{g}_{\mbf{u}^\star}( \mbf{i})$ denote such an approximation, defined as:
\[
\overline{g}_{\mbf{u}^\star}( \mbf{i}) = \mbf{u}^\star + \bm{\delta} - \bm{\delta} \odot \left( \mbf{i} \oslash \mbf{t} \right).
\]
Note that when the incentive is $\mbf{0}$, $\overline{g}_{\mbf{u}^\star}( \mbf{0}) = g_{\mbf{u}^\star}(\mbf{0})$, and when the incentive is $\mbf{t}$, $\overline{g}_{\mbf{u}^\star}( \mbf{t}) = g_{\mbf{u}^\star}(\mbf{t}) = \mbf{u}^\star$.  In the intermediate region, $\overline{g}$ is a simple linear interpolation.
\end{dfn}

\section{Feedback Optimization Algorithms}
\label{sec:opt}

In this section, we present feedback-based optimization algorithms designed to solve \eqref{eq:sci} from the perspective of the system operator.  Recall that this setting assumes that the function $g_{\mbf{u}^\star}( \cdot )$is unknown to the SO, but specific outcomes such as $\mbf{u} = g_{\mbf{u}^\star}(\mbf{i}')$ and $\nabla_{\mbf{i}} g_{\mbf{u}^\star}(\mbf{i}')$ for a given incentive $\mbf{i}'$ can be \textit{measured} or \textit{estimated}.

We start with a \textit{warmup algorithm} that we henceforth term ''iterative incentive increase'' (\texttt{III}).  This heuristic approach starts with a zero incentive (i.e., $\mbf{i}^{(0)} = \mbf{0}$), and increases from there based on the \textit{distance} between the operating point $\mbf{u}' = g_{\mbf{u}^\star}(\mbf{i}^{(k)})$ and the setpoint $\mbf{u}^\star$.  By the monotonicity of $g_{\mbf{u}^\star}( \cdot )$ (see \sref{Assumption}{asm:monotone}), if $\mbf{u}$ is \textit{larger than} $\mbf{u}^\star$, increasing the incentive will bring $\mbf{u}$ towards $\mbf{u}^\star$, and vice versa. Below, in \autoref{alg:iii}, we formalize this idea, using a hyperparameter $\epsilon > 0$ as the multiplicative step size.
\begin{algorithm}[!h]
    \small
	\caption{Iterative Incentive Increase (\texttt{III})}
	\label{alg:iii}
	\begin{algorithmic}[1]
		\State \textbf{initialize:} Initialize $\mbf{i}^{(0)} = \mbf{0}$ (no incentive) and step size $\epsilon > 0$.
		\For{steps $k = 1, 2, \dots$}
            \State Measure $g_{\mbf{u}^\star}(\mbf{i}^{(k)})$;
            \State Compute $\mbf{r}^{(k)} \leftarrow \left( g_{\mbf{u}^\star}(\mbf{i}^{(k)}) - \mbf{u}^\star \right)$ elementwise, preserve sign;
            \State Update incentive: $\mbf{i}^{(k)} + \epsilon  \mbf{r}^{(k)}$ %
        \EndFor
	\end{algorithmic}
\end{algorithm}

For sufficiently small $\epsilon$, it is straightforward to observe that \texttt{III} converges at the point where $g_{\mbf{u}^\star}(\mbf{i}) = \mbf{u}^\star$; while feasible, this point is likely \textit{suboptimal} in terms of $c(\mbf{i})$.  Thus, although \texttt{III} will not perform optimally, it is easy to implement and will serve as a useful baseline %
in our numerical experiments (see \autoref{sec:exp}).

To attain the (near-)optimal incentive when $g_{\mbf{u}^\star}( \cdot )$ is unknown to the SO, in the following sections we present three iterative feedback algorithms that require varying amounts of information about the underlying problem.  We start with a dual ascent technique that achieves the optimal incentive when given significant information about $g_{\mbf{u}^\star}( \cdot )$ before considering techniques that require less information but achieve weaker theoretical guarantees.

\subsection{Finding the Optimum Using Dual Ascent}
Since \eqref{eq:sci} is \textit{constrained}, a natural place to start is by considering a \textit{dual ascent} iterative algorithm.  To do so, we introduce the Lagrangian associated with \eqref{eq:sci}:
\begin{equation}
\mathcal{L}(\mbf{i}, \bm{\lambda}) = c(\mbf{i}) + \bm{\lambda}^\top (h(g_{\mbf{u}^\star}(\mbf{i}))), \quad \quad \bm{\lambda} \geq \mbf{0}, \label{eq:lagrang}
\end{equation}
where the vector $\bm{\lambda}$ represents dual variables.  Given this Lagrangian, $\DAIO$ (Dual Ascent Incentive Optimization) is given by \autoref{alg:daio}.
\begin{algorithm}[!h]
    \small
	\caption{Dual Ascent Incentive Optimization (\texttt{DAIO})}
	\label{alg:daio}
	\begin{algorithmic}[1]
        \State \textbf{input:} Hyperparameter $\epsilon > 0$.
		\State \textbf{initialize:} Initialize $\mbf{i}^{(0)} \succcurlyeq \mbf{0}$ and initial dual guess $\bm{\lambda}^{(0)} \succ \mbf{0}$.
		\For{steps $k = 1, 2, \dots$}
            \State Update incentive: $\mbf{i}^{(k)} \in \argmin_{\mbf{i}} \mathcal{L}(\mbf{i}, \bm{\lambda}^{(k-1)})$;
            \State Dual update: $\bm{\lambda}^{(k)} = \left[ \bm{\lambda}^{(k-1)} + \epsilon \cdot h(g_{\mbf{u}^\star}(\mbf{i}^{(k)})) \right]_{\mathbb{R}^n_{+}}$.
        \EndFor
	\end{algorithmic}
\end{algorithm}\vspace{-1em}

\noindent When the \textit{functional form} of $g_{\mbf{u}^\star}(\mbf{i})$ and the gradient $\nabla_{\mbf{i}} g_{\mbf{u}^\star}(\mbf{i})$ are both known, we can derive a closed form for the $\argmin$ term on line 4.  The input to the dual update step on line 5 comes from measurements of $h(g_{\mbf{u}^\star}(\mbf{i}^{(k)}))$, which are violations of the safety constraint.

\subsubsection{Closed-form for quadratic-convex $g_{\mbf{u}^\star}(\mbf{i})$} \label{sec:quad-convex-CF-DAIO}
Suppose that $g_{\mbf{u}^\star}(\mbf{i})$ adopts a functional form that is quadratic and convex in $\mbf{i}$.  In particular, if $\bm{\delta}$ represents the deviance from $\mbf{u}^\star$ when $\mbf{i} = \mbf{0}$ and $\mbf{t}$ is the threshold vector from \sref{Assumption}{asm:threshold}, we let $\mbf{b} = \bm{\delta} \oslash \left( \mbf{t}^{\circ 2} \right)$, and assume that 
\[
g_{\mbf{u}^\star}(\mbf{i}) = \left[ u^\star_{0} + b_{0} (i_{0} - t_{0})^2, \quad \dots, %
\quad u^\star_{n} + b_{n} ( i_{n} - t_{n} )^2 \right].
\]
Since $g_{\mbf{u}^\star}(\mbf{i})$ is separable across dimensions%
, we can express $\nabla_{\mbf{i}} g_{\mbf{u}^\star}(\mbf{i})$ in the compact form $
\nabla_{\mbf{i}} g_{\mbf{u}^\star}(\mbf{i}) = \text{diag}(2\mbf{b}) \text{diag}(\mbf{i}) - \text{diag}(2\mbf{b}) \text{diag}(\mbf{t})$.
The $\argmin$ term on line 4 of \autoref{alg:daio} simplifies as follows:
\begin{align}
    \mbf{i}^{(k)} &\in \argmin_{\mbf{i}} \mathcal{L}(\mbf{i}, \bm{\lambda}^{(k-1)})\\
    &= \begin{cases}
        \mbf{0} \quad &\text{if } \bm{\lambda} = \mbf{0},\\
        \argmin_{\mbf{i}} c(\mbf{i}) + \bm{\lambda}^\top h(g_{\mbf{u}^\star}(\mbf{i})) \quad &\text{if } \bm{\lambda} > \mbf{0}.
    \end{cases} \label{eq:argminCF}
\end{align}
Furthermore, recall that the incentive $\mbf{i}$ that minimizes \eqref{eq:argminCF} satisfies the Lagrangian stationarity condition $\mbf{0} \in \partial_{\mbf{i}} \left[ c(\mbf{i}) + \bm{\lambda}^\top h(g_{\mbf{u}^\star}(\mbf{i})) \right],$
where $\partial$ denotes subgradients w.r.t. $\mbf{i}$ of $\mathcal{L}(\mbf{i}, \bm{\lambda})$.  Noting that $h(g_{\mbf{u}^\star}(\mbf{i}))$ is differentiable everywhere, we have:
\begin{equation*}
\mbf{0} %
\in \partial_{\mbf{i}} c(\mbf{i}) + \nabla_{\mbf{i}} g_{\mbf{u}^\star}(\mbf{i})^\top \nabla h(g_{\mbf{u}^\star}(\mbf{i}))  \bm{\lambda},
\end{equation*}
For an application to distribution grid voltage control (see \autoref{sec:distgridVC}), we let $c(\mbf{i}) \coloneqq \Vert \mbf{i} \Vert_1$, which implies that $\mbf{1} \in \partial_{\mbf{i}} c(\mbf{i})$.  Under assumptions that only the voltage lower bound is a safety concern, and that only active power $\mbf{p}$ is sensitive to incentives, we consider the simpler form of $h(\cdot)$ given in \eqref{eq:simple-h}.  This gives the following compact simplification:
\begin{equation}
\mbf{0} = \mbf{1} - \nabla_{\mbf{i}} g_{\mbf{u}^\star}(\mbf{i})^\top \mbf{R}^\top \bm{\lambda} \label{eq:subgLagrang}
\end{equation}
Combined with the quadratic-convex functional form of $g$, this then yields the following:
\begin{align*}
-\mbf{1} &= - \left( \text{diag}(2\mbf{b}) \text{diag}(\mbf{i}) - \text{diag}(2\mbf{b}) \text{diag}(\mbf{t}) \right)^\top \mbf{R}^\top \bm{\lambda},\\
-\mbf{1} &= - \text{diag}(\mbf{i}) \text{diag}(2\mbf{b}) \mbf{R}^\top \bm{\lambda} + \text{diag}(\mbf{t}) \text{diag}(2\mbf{b}) \mbf{R}^\top \bm{\lambda},\\
\text{diag}(\mbf{i}) \text{diag}(2\mbf{b}) \mbf{R}^\top \bm{\lambda} &= \text{diag}(\mbf{t}) \text{diag}(2\mbf{b}) \mbf{R}^\top \bm{\lambda} + \mbf{1},\\
\mbf{i} &= \mbf{t} + \left( \mbf{R}^\top (2\mbf{b} \odot \bm{\lambda}) \right)^{\circ -1}.
\end{align*}
where in the last step, the Hadamard inverse ($^{\circ -1}$) indicates taking the reciprocal of each element.
Thus, when $g$ is known to adopt this quadratic-convex form, the $\argmin$ term on line 4 in \autoref{alg:daio} can be replaced with the following:
\begin{equation}
    \mbf{i}^{(k)} = \mbf{t} + \left( \mbf{R}^\top (2\mbf{b} \odot \bm{\lambda}^{(k-1)}) \right)^{\circ -1} \in \argmin_{\mbf{i}} \mathcal{L}(\mbf{i}, \bm{\lambda}^{(k-1)}) \label{eq:closedformPrimalUpdateDAIO}
\end{equation}

\subsubsection{Convergence properties} 
In the following, we show that \DAIO converges to the unique minimizer of \eqref{eq:sci} under mild assumptions.  We note that this result does not require a closed form for the $\argmin$ incentive update (as derived in \autoref{sec:quad-convex-CF-DAIO}).  We start with the necessary assumption before stating the result.

\begin{ass} \label{asm:slater}
\eqref{eq:sci} is convex and satisfies Slater's condition.
\end{ass}

Recall that Slater's condition implies that \textit{strong duality} holds for Problem \eqref{eq:sci}.  In the following, we show that when the step size $\epsilon$ satisfies an inequality, \DAIO converges to the unique minimizer under the above assumptions.

\begin{thm} \label{thm:DAIOconvergence}
Under \sref{Assumption}{asm:slater}, \DAIO (\autoref{alg:daio}) converges to the unique minimizer of \eqref{eq:sci} when
\begin{equation}
    0 < \epsilon < \frac{2 m(\bm{\lambda}^{(0)})}{\Vert h(g_{\mbf{u}^\star}(\mbf{i}^{(0)})) \Vert_2^2}, \label{eq:epsCondition}
\end{equation}
where $m$ is the dual problem of \eqref{eq:sci}, i.e., $m(\bm{\lambda}) = \min_{\mbf{i}} \mathcal{L}(\mbf{i}, \bm{\lambda})$.
\end{thm}

We defer the proof of \autoref{thm:DAIOconvergence} to \sref{Appendix}{apx:DAIOconvergence}.
Although this result shows that we can derive an effective iterative algorithm to find the optimal solution when properties of $g_{\mbf{u}^\star}(\mbf{i})$ are known, in the following section we consider the case where the SO only has access to (estimates of) the gradient $\nabla_{\mbf{i}} g_{\mbf{u}^\star}(\mbf{i})$.

\subsection{Leveraging First-Order Gradient Information}
We consider the case where the SO has knowledge (or feedback-based estimates) of $\nabla_{\mbf{i}} g_{\mbf{u}^\star}(\mbf{i})$, but otherwise does not have any other information about $g_{\mbf{u}^\star}(\mbf{i})$ as required in the previous section.  We present a primal-dual gradient-based method that we henceforth denote by \FOIO (First-Order Incentive Optimization, \autoref{alg:foio}).

Recall the Lagrangian given in \eqref{eq:lagrang}.  In \autoref{sec:quad-convex-CF-DAIO}, \eqref{eq:subgLagrang} implicitly derives the (sub)gradient of the Lagrangian for the simpler form of $h(\cdot)$ where only the voltage lower bound is a concern.  The subgradient reads as follows:
\begin{equation*}
\nabla_{\mbf{i}} \mathcal{L}(\mbf{i}^{(k)}, \bm{\lambda}^{(k)}) \coloneqq \mathbf{1} - \nabla_{\mbf{i}} g_{\mbf{u}^\star}(\mbf{i}^{(k)})^\top \mathbf{R}^\top \bm{\lambda}^{(k)}
\end{equation*}
The primal-dual gradient-based approach is given by \autoref{alg:foio}.
Note that compared to \DAIO, \FOIO follows a broad literature on gradient-based methods~\cite{BV:04, Bertsekas:09, Nesterov:13} %
and introduces a time-varying step size $\epsilon_k$.  In what follows, we characterize the ``correct choice'' of  $\{ \epsilon_k > 0 \}_{k \geq 1}$ that facilitates convergence.

\begin{algorithm}[h]
    \small
	\caption{First-Order Incentive Optimization (\texttt{FOIO})}
	\label{alg:foio}
	\begin{algorithmic}[1]
        \State \textbf{input:} Time-varying step sizes $\{ \epsilon_k > 0 \}_{k \geq 1}$.
		\State \textbf{initialize:} Initialize primal and dual guesses $\mbf{i}^{(0)}, \bm{\lambda}^{(0)} \succcurlyeq \mbf{0}$..
		\For{steps $k = 1, 2, \dots$}
            \State Incentive (primal) update: $\mbf{i}^{(k)} = \mbf{i}^{(k-1)} - \epsilon_k \nabla_{\mbf{i}} \mathcal{L}(\mbf{i}^{(k)}, \bm{\lambda}^{(k)})$;
            \State Dual update: $\bm{\lambda}^{(k)} = \left[ \bm{\lambda}^{(k-1)} + \epsilon_k \cdot h(g_{\mbf{u}^\star}(\mbf{i}^{(k)})) \right]_{\mathbb{R}^n_{+}}$.
        \EndFor
	\end{algorithmic}
\end{algorithm}

\subsubsection{Convergence analysis}
In the following, we show that $\FOIO$ (First-Order Incentive Optimization) converges to the unique minimizer of \eqref{eq:sci} using a technique given by~\cite{Boyd:14}.  We continue to assume \sref{Assumption}{asm:slater} holds.  Furthermore, we assume:

\begin{ass} \label{asm:boundedNorms}
The (Euclidean) norms of the (sub)gradients of $c(\mathbf{i})$ and $h(g_{\mbf{u}^\star}(\mbf{i}))$ are finite for any incentive $\mbf{i}$.
\end{ass}

Below, we show that when the time-varying step sizes $\{ \epsilon_k \}_{k \geq 1}$ satisfy certain conditions, \FOIO converges to the unique minimizer under the above assumptions. 

\begin{thm} \label{thm:FOIO-convergence}
Under \sref{Assumptions}{asm:slater} and \ref{asm:boundedNorms}, \FOIO (\autoref{alg:foio}) converges to the unique minimizer of \eqref{eq:sci} if the step sizes are defined as
\begin{equation}
    \epsilon_k = \frac{\gamma_k}{\Vert \bm{\psi}^{(k)} \Vert_2}, \label{eq:epsConditionFO}
\end{equation}
where $\gamma_k$ is a square summable but not summable positive quantity, and $\bm{\psi}^{(k)}$ collects the (sub)gradients of $\mathcal{L}$ with respect to $i$ and $\bm{\lambda}$.
\end{thm}

We defer the proof of \autoref{thm:FOIO-convergence} to \sref{Appendix}{apx:FOIO-convergence}.
Although this result shows that \FOIO finds the optimum when the exact gradient $\nabla_{\mbf{i}} g_{\mbf{u}^\star}(\mbf{i})$ is known, in the following we also consider the case where \FOIO must contend with estimates of this quantity.

\subsubsection{Convergence analysis with inaccurate gradients}
While the previous analysis gives strong guarantees when \FOIO has access to the gradient $\nabla g_{\mbf{u}^\star}(\mbf{i})$, recall that since $g_{\mbf{u}^\star}(\mbf{i})$ is unknown to the SO, it is likely that only estimates of this sensitivity are available in practice.  In what follows, we prove stability and convergence properties when \FOIO is provided with coarse estimates of $\nabla_{\mbf{i}} g$.

Recall \sref{Assumptions}{asm:monotone}, \ref{asm:threshold}, and \ref{asm:slater}.  For the purpose of the following analysis, we assume that the true $g_{\mbf{u}^\star}(\mbf{i})$ is equal to the linear $\overline{g}$ given in \sref{Definition}{dfn:linear-g}.  Recall that $\overline{g}_{\mbf{u}^\star}(\mbf{i})$ adopts the following (linear) form, where $\bm{\delta}$ represents the deviance from $\mbf{u}^\star$ when $\mbf{i} = \mbf{0}$.
\[
\overline{g}_{\mbf{u}^\star}(\mbf{i}) = \left[ u^\star_{0} + \delta_{0} - \delta_{0} \frac{i_{0}}{t_{0}}, \quad \dots, %
\quad u^\star_{n} + \delta_{n} - \delta_{n} \frac{i_{n}}{t_{n}} \right]
\]
Note that when $g_{\mbf{u}^\star}(\mbf{i}) = \overline{g}_{\mbf{u}^\star}( \mbf{i})$, we have $\nabla_{\mbf{i}} g_{\mbf{u}^\star}(\mbf{i}) = \text{diag}( -\bm{\delta} \oslash \mbf{t} )$.

With respect to a linear $g$ function as outlined above, suppose that \FOIO receives a \textit{coarse estimate} of $\nabla_{\mbf{i}} g$.  First, we let $\mbf{t'}$ denote a vector that estimates the true value of $\mbf{t}$ across all buses.
Then suppose that \FOIO is given a gradient estimate $\widehat{\nabla}_i g_{\mbf{u}^\star}(\mbf{i}) = \text{diag}( -\bm{\delta} \oslash \mbf{t'} )$.  Here we note that $\bm{\delta}$ (i.e., the deviance from the setpoint $\mbf{u}^\star$ when the incentive is $\mbf{0}$) can be measured, and an estimate of  $\mbf{t'}$ may be obtained from e.g., a linear regression on prior data points.

To show \textit{error bounds} on the solution obtained by \FOIO (with respect to an underlying linear $\overline{g}_{\mbf{u}^\star}( \mbf{i})$) when it is given a coarse estimate as outlined above, we start with some assumptions on Problem \eqref{eq:sci}.

\begin{ass} \label{asm:lipschitz-objective}
 $c(\mbf{i})$ is convex and continuously differentiable.  The (sub)gradients $\partial c(\mbf{i})$ and $\partial^2 c(\mbf{i})$ are Lipschitz continuous.
\end{ass}
\begin{ass} \label{asm:lipschitz-constraint}
    The composition $h(g_{\mbf{u}^\star}(\mbf{i}))$ is convex and continuously differentiable, and the gradients $\nabla h(g_{\mbf{u}^\star}(\mbf{i}))$ and $\nabla^2 h(g_{\mbf{u}^\star}(\mbf{i}))$ are Lipschitz continuous.
\end{ass}
\begin{thm} \label{thm:FOIO-convergence-bad-T}
Under \sref{Assumptions}{asm:slater}, \ref{asm:lipschitz-objective}, and \ref{asm:lipschitz-constraint}, \FOIO's iterates with coarse gradient estimates $\widehat{\nabla}_i g_{\mbf{u}^\star}(\mbf{i})$ satisfy
\begin{equation}
\lim_{k \to \infty} \sup \Vert \mbf{i}^{(k)} - \mbf{i}^\star \Vert_2 = O(\epsilon \Vert ( \bm{\delta} \oslash \mbf{t'} ) - ( \bm{\delta} \oslash \mbf{t} ) \Vert_2 ), \label{eq:FOinexactbound}
\end{equation}
where $\mbf{i}^\star$ is the optimal solution to \eqref{eq:sci} when $g_{\mbf{u}^\star}(\mbf{i}) = \overline{g}_{\mbf{u}^\star}( \mbf{i})$.
\end{thm}

We defer the proof of \autoref{thm:FOIO-convergence-bad-T} to \sref{Appendix}{apx:FOIO-convergence-bad-T}.
The result implies that even when \FOIO is given only coarse estimates of the gradient $\nabla_i g_{\mbf{u}^\star}(\mbf{i})$, it will still converge to within a reasonable margin of the optimal solution -- we explore this experimentally in \autoref{sec:exp-results-station}, finding complementary results.

\subsection{Model-Free Zero-Order Optimization for Unknown Environments}
In practice, if $g_{\mbf{u}^\star}(\mbf{i})$ is truly unknown to the SO, it is also reasonable to consider the case where the SO  does not have knowledge (or even estimates) of the sensitivity matrix $\nabla_i g_{\mbf{u}^\star}(\mbf{i})$ required for \FOIO.  To handle this case, we follow related work \cite{Cavraro:24, Cui:23, Chen:20} to derive a \textit{model-free zero-order method} that estimates the gradient of the Lagrangian using two function evaluations.

\noindent Recall that the Lagrangian $\mathcal{L}(\mbf{i}, \bm{\lambda})$ is given by \eqref{eq:lagrang}. We introduce the following \textit{regularized Lagrangian}~\cite{Koshal:11}:
\begin{equation}
\mathcal{L}_{p,d}(\mbf{i}, \bm{\lambda}) = \mathcal{L}(\mbf{i}, \bm{\lambda}) + \frac{p}{2} \Vert \mbf{i} \Vert_2^2 - \frac{d}{2} \Vert \bm{\lambda} \Vert_2^2, \label{eq:reg-lagrang}
\end{equation}
where $p > 0$ and $d > 0$ are regularization parameters.  Under this regularization, \eqref{eq:sci} can be expressed as the saddle-point problem:
\begin{equation}
    \max_{\bm{\lambda} \in \mathbb{R}^n_+} \min_{\mbf{i} \in \mathbb{R}^n} \mathcal{L}_{p,d}(\mbf{i}, \bm{\lambda}). \label{eq:sci-reg}
\end{equation}
Note that $\mathcal{L}_{p,d}(\mbf{i}, \bm{\lambda}) $ is strongly convex in $\mbf{i}$ and strongly concave in $\bm{\lambda}$, implying that the optimal solution $\mbf{z}^\star = (\mbf{i}^\star, \bm{\lambda}^\star)$ associated with $\mathcal{L}_{p,d}$ is unique.  However, $\mbf{z}^\star$ may be different from the saddle points of the exact Lagrangian \eqref{eq:lagrang} -- bounds on the distance between $\mbf{z}^\star$ and the solution of \eqref{eq:sci} can be established~\cite{Koshal:11}.

We then consider a model-free two function evaluation approximation of the Lagrangian gradient for a given iteration $k \in \mathbb{N}$~\cite{Chen:20}:
\begin{equation}
    \widehat{\nabla} \mathcal{L}^{(k)} \coloneqq \frac{\bm{\zeta}^{(k)}}{2\sigma} \left[ \hat{\mathcal{L}}(\mbf{i}_{+}^{(k)}, \bm{\lambda}^{(k)}) - \hat{\mathcal{L}}(\mbf{i}_{-}^{(k)}, \bm{\lambda}^{(k)}) \right], \label{eq:ZOapprox}
\end{equation}
where $\mbf{i}_{\pm}^{(k)} \coloneqq \mbf{i}^{(k)} \pm \sigma \bm{\zeta}^{(k)}$ denote \textit{perturbed incentives}.  Here $\sigma > 0$ is a parameter that controls the magnitude of perturbation, and $\bm{\zeta}^{(k)} \in \mathbb{R}^n$ is a random perturbation signal.  The $\hat{\mathcal{L}}$ notation indicates that the value of the Lagrangian for a given perturbed incentive may be e.g., quickly measured and is susceptible to error.
Using this approximation of the gradient, the algorithm (\ZOIO, Zero-Order Incentive Optimization) is given by \autoref{alg:zoio}.
\begin{algorithm}[!h]
\small
	\caption{Zero-Order Incentive Optimization (\texttt{ZOIO})}
	\label{alg:zoio}
	\begin{algorithmic}[1]
        \State \textbf{input:} Hyperparameters $\sigma, \epsilon > 0$.
		\State \textbf{initialize:} Initialize primal and dual guesses $\mbf{i}^{(0)}, \bm{\lambda}^{(0)} \succcurlyeq \mbf{0}$.
		\For{steps $k = 1, 2, \dots$}
            \State Estimate Lagrangian gradient $\widehat{\nabla} \mathcal{L}^{(k-1)}$ according to \eqref{eq:ZOapprox}
            \State Incentive (primal) update: $\mbf{i}^{(k)} = (1 - \epsilon p) \mbf{i}^{(k-1)} - \epsilon \widehat{\nabla} \mathcal{L}^{(k-1)}$;
            \State Dual update: $\bm{\lambda}^{(k)} = \left[ (1 - \epsilon d) \bm{\lambda}^{(k-1)} + \epsilon h(g_{\mbf{u}^\star}(\mbf{i}^{(k)})) \right]_{\mathbb{R}^n_{+}}$.
        \EndFor
	\end{algorithmic}
\end{algorithm}

Compared to the previous algorithms \DAIO and \FOIO, the primary benefit of \ZOIO is that it can be implemented in complete model-free fashion, provided that measurements are available of $g_{\mbf{u}^\star}( \mbf{i'})$ for a given incentive $\mbf{i'}$.  Furthermore, in what follows, we can show guarantees on the asymptotic distance between \ZOIO's incentives and the optimal solution under some natural assumptions.

\subsubsection{Convergence to near-optimal points}
First, recall \sref{Assumptions}{asm:slater}, \ref{asm:lipschitz-objective}, and \ref{asm:lipschitz-constraint}.  We further assume that the \textit{measurement error} associated with $\hat{\mathcal{L}}$ is bounded as follows:
\begin{ass} \label{asm:measurement-error}
    There exists a finite scalar $e_y$ such that for all $k$ and $\mbf{i}$, the \textit{measurement error} of the operating point $\mbf{u}$ (given by $g_{\mbf{u}^\star}( \cdot )$) can be bounded as:
    \begin{align*}
        \sup_{k \ge 1} \Vert \hat{g}^{(k)}_{-} - g_{\mbf{u}^\star}(\mbf{i}^{(k)}_{-}) \Vert \leq e_y,\\
        \sup_{k \ge 1} \Vert \hat{g}^{(k)}_{+} - g_{\mbf{u}^\star}(\mbf{i}^{(k)}_{+}) \Vert \leq e_y.
    \end{align*}
\end{ass}
Since $c(\cdot)$ and $h(\cdot)$ are known to the SO, \sref{Assumption}{asm:measurement-error} immediately gives a corresponding error bound for $\hat{\mathcal{L}}$.
We start by recalling \cite[Lemma 1]{Chen:20}, which states the following: for a $C^3$ function $\mathcal{L} : \mathbb{R}^n \to \mathbb{R}$ with Lipschitz continuous $\nabla \mathcal{L}$ and $\nabla^2 \mathcal{L}$ that is approximated using two function evaluations as in \eqref{eq:ZOapprox}, we have 
\[
\widehat{\nabla} \mathcal{L} = \bm{\zeta} \bm{\zeta}^\top \nabla \mathcal{L}(\mbf{i}) + O(\sigma^2),
\]
where $\nabla \mathcal{L}(\mbf{i})$ slightly abuses notation to indicate $\nabla_{\mbf{i}} \mathcal{L} (\mbf{u}, \mbf{i})$.
To apply \cite[Lemma 1]{Chen:20} for the purposes of convergence, we will assume that $\bm{\zeta}$ is an exploration signal that satisfies the following:
\begin{ass} \label{asm:ident}
    The exploration signal $\bm{\zeta}$ is a periodic signal with period $P$, and for all $t$:
    \begin{equation}
    \frac{1}{P} \int_{t}^{t+P} \bm{\zeta}(\tau) \bm{\zeta}(\tau)^\top d\tau = \mathbf{I}. \label{eq:zetaIdent}
    \end{equation}
\end{ass}
An example of an exploration signal that satisfies the above assumption is a sinusoidal signal with element-wise unique frequencies given by \cite[(16)]{Chen:20}.
In what follows, we show that the primal update in \ZOIO (\autoref{alg:zoio}, line 5) is approximately equivalent to an averaged primal step, where the ``gain matrix'' $\bm{\zeta} \bm{\zeta} ^\top$ in the approximation is replaced with the identity matrix \eqref{eq:zetaIdent}.  
\begin{thm} \label{thm:ZOconvergence-nontimevary}
Under \sref{Assumptions}{asm:slater}, \ref{asm:lipschitz-objective}, \ref{asm:lipschitz-constraint}, \ref{asm:measurement-error}, and \ref{asm:ident}, for a continuous time model, \ZOIO's iterates satisfy
\begin{equation}
\lim_{\tau \to \infty} \sup \Vert \mbf{i}(\tau) - \mbf{i}^\star \Vert = O(\epsilon + \sigma^2 + e_y). \label{eq:ZObound}
\end{equation}
\end{thm}

We defer the proof of \autoref{thm:ZOconvergence-nontimevary} to \sref{Appendix}{apx:ZOconvergence-nontimevary}.  The proof considers the algorithm's primal and dual updates in \textit{continuous time}, using ODE approximation techniques from stochastic approximation literature.  Using the periodicity of $\bm{\zeta}$, bounds on the aggregate updates yield a uniform bound on the distance between \ZOIO and the optimum that is intuitively a function of the error and variability in the zero-order gradient approximations.

\section{Non-Stationary Environment}
\label{sec:timevarying}
In this section, we formalize a light extension of the formulation given in \autoref{sec:prob} that considers a \textit{non-stationary environment}.  In particular, motivated by the intended application to distribution grid control, we consider the case where the $g_{\mbf{u}^\star}(\mbf{i})$ function is allowed to be \textit{time-varying} -- this captures a case where both the underlying demand and the underlying preferences (e.g., of end-users) do not remain constant during the optimization process.

We start by formalizing the problem as follows, where $k \in \mathbb{N}$ denotes the \textit{time index} of the problem:
\begin{align}
&\begin{aligned}
\min_{ \{ \mbf{i}^{(k)} \in \mathbb{R}^n \}_{k \ge 1}} & \ \ \sum_{k \ge 1} c(\mbf{i}^{(k)}),  \quad (\textit{cost of incentives over time})\\
\text{s.t., } & \ \ h \left( g^{(k)}_{\mbf{u}^{(\star,k)}}(\mbf{i}^{(k)}) \right) \preccurlyeq \mbf{0} \ \ \forall k \in \{1, 2, \dots\}
\end{aligned} \label{eq:sci-timevarying}
\end{align}

We assume that the SO specifies a setpoint $\mbf{u}^{(\star,k)}$ that may or may not be fixed in $k$.  At each time step $k$, they incentivize end-users according to a time-varying incentive given by $\mbf{i}^{(k)}$, where we note that the optimal incentive may change over time (and is henceforth denoted by $\mbf{i}^{(\star, k)}$). The cost function $c(\cdot)$ remains constant across all time steps.
The constraint function $h(\cdot)$ is assumed to be constant and known across all time steps $k$, but the environment's response function $g^{(k)}_{\mbf{u}^{(\star,k)}}(\mbf{i})$ varies with time and remains unknown across all time steps.  Although $g^{(k)}_{\mbf{u}^{(\star,k)}}( \cdot )$ is unknown, we assume the SO can measure (sample) outcomes.

The Lagrangian associated with \eqref{eq:sci-timevarying} is given by the following:
\begin{equation}
\mathcal{L}^{(k)}(\mbf{i}^{(k)}, \bm{\lambda}^{(k)}) = c(\mbf{i}^{(k)}) + \bm{\lambda}^{(k) \top} (h(g^{(k)}_{\mbf{u}^{(\star,k)}}(\mbf{i}^{(k)}))), \quad \quad \bm{\lambda}^{(k)} \geq \mbf{0}, \label{eq:lagrang-timevarying}
\end{equation}
where note that, like the incentive $\mbf{i}^{(k)}$, the dual variables $\bm{\lambda}^{(k)}$ and optimal dual variables $\bm{\lambda}^{(\star, k)}$ are time-varying.  Similarly, we define the following \textit{regularized} Lagrangian for \ZOIO below:
\begin{equation}
\mathcal{L}_{p,d}^{(k)}(\mbf{i}^{(k)}, \bm{\lambda}^{(k)}) = \mathcal{L}^{(k)}(\mbf{i}^{(k)}, \bm{\lambda}^{(k)}) + \frac{p}{2} \Vert \mbf{i}^{(k)} \Vert_2^2 - \frac{d}{2} \Vert \bm{\lambda}^{(k)} \Vert_2^2, \label{eq:reg-lagrang-timevarying}
\end{equation}
where the regularization parameters $p, d > 0$ are \textit{not} time-varying.

Under a few small tweaks, the feedback-based optimization algorithms presented in the previous section naturally extend to this non-stationary environment.  Namely, the incentive update and dual variable updates in \FOIO and \ZOIO must be slightly tweaked to swap the non-time-varying Lagrangian \eqref{eq:lagrang} for the time-varying Lagrangian \eqref{eq:lagrang-timevarying}.  To avoid the ``vanishing gradient'' phenomenon, in this time-varying case we also set a fixed step size $\epsilon > 0$ in \FOIO (\autoref{alg:foio}, line 1), such that $\epsilon_k = \epsilon : k \geq 1$.

We continue by stating theoretical results on the \textit{tracking} properties of these algorithms under appropriate assumptions -- these give provable bounds on the asymptotic gap between the iterates of our algorithms and the time-varying optimal incentive.
For the purposes of the following analysis, we recall \sref{Assumptions}{asm:slater}, \ref{asm:lipschitz-objective}, and \ref{asm:lipschitz-constraint}, further assuming that each holds across all steps $k$.  Recalling that $\mbf{z}^{(k)} = (\mbf{i}^{(k)}, \bm{\lambda}^{(k)})$, we quantify the temporal variability of the optimal incentive and dual variables by $\omega^{(k)}$ as follows:
\begin{equation}
\omega^{(k)} \coloneqq \Vert \mbf{z}^{(\star, k)} - \mbf{z}^{(\star, k-1)} \Vert. \label{eq:sigmadef}
\end{equation}

\subsubsection{Tracking properties of \FOIO}

We first define a set-valued and time-varying mapping $\bm{\psi}^{(k)}$ that collects the primal and dual operators corresponding to the iterates of \FOIO:
\begin{align}
\bm{\psi}^{(k)}(\mbf{i}^{(k)},\bm{\lambda}^{(k)}) = \begin{bmatrix}
    \partial_{\mbf{i}} \mathcal{L}^{(k)}(\mbf{i}^{(k)}, \bm{\lambda}^{(k)}) \\
    - h(g^{(k)}_{\mbf{u}^{(\star,k)}}( \mbf{i}^{(k)}))
\end{bmatrix}.
\end{align}

Under the conditions of \sref{Assumptions}{asm:slater}, \ref{asm:lipschitz-objective}, and \ref{asm:lipschitz-constraint}, the following Lemma holds, as shown in~\cite[Lemma 5]{Bernstein:19}:

\begin{lem} \label{lem:psi-monotone}
There exist finite constants $0 < \eta_{\bm{\psi}} < L_{\bm{\psi}}$ such that for all $k$, the map $\bm{\psi}^{(k)}$ is $\eta_{\bm{\psi}}$-strongly monotone over $\mathbb{R}^n \times \mathbb{R}^n_+$ and $L_{\bm{\psi}}$-Lipschitz over $\mathbb{R}^n \times \mathbb{R}^n_+$.
\end{lem}

Under \sref{Lemma}{lem:psi-monotone} and \cite[Theorem 4]{Bernstein:19}, it immediately follows that \FOIO exhibits the following tracking performance. 

\begin{theorem} \label{thm:FOIO-convergence-timevarying}
Assume that there exists a finite scalar $\omega$ such that $\sup_{k \ge 1} \omega^{(k)} \le \omega$, and let the (non-time-varying) step size $\epsilon$ be chosen such that $0 < \epsilon < \nicefrac{2\eta_{\bm{\psi}}}{L_{\bm{\psi}}^2}$.  Then the sequence of iterates $\{ \mbf{z}^{(k)} \}$ converges Q-linearly to  $\{ \mbf{z}^{(\star, k)} \}$ up to an asymptotic error bound given by
\begin{equation}
\lim_{k \to \infty} \sup \Vert \mbf{z}^{(k)} - \mbf{z}^{(\star, k)} \Vert_2 \leq \frac{\omega}{1-\sqrt{1- 2\epsilon \eta_{\bm{\psi}} + \epsilon^2 L_{\bm{\psi}}^2}}.
\end{equation}
\end{theorem}
We note that this error bound intuitively depends on the temporal variability of the optimal solution ($\omega$), and the properties of $\bm{\psi}$, namely monotonicity ($\eta_{\bm{\psi}}$) and Lipschitzness ($L_{\bm{\psi}}$).

\subsubsection{Tracking Properties of \ZOIO}
To prove tracking properties for \ZOIO, we additionally recall the relevant zero-order assumptions, namely \sref{Assumptions}{asm:measurement-error} and \ref{asm:ident}. We then further assume that the temporal variability of the gradients $\nabla h(g^{(k)}_{\mbf{u}^{(\star,k)}}(\mbf{i}))$ satisfies:
\begin{ass} \label{asm:variability-gradient}
    There exists a finite constant $e_f$ such that for all $k$ and $\mbf{i}$:
    \[
    \Vert \nabla h(g^{(k)}_{\mbf{u}^{(\star,k)}}(\mbf{i})) - \nabla h(g^{(k-1)}_{\mbf{u}^{(\star,k-1)}}(\mbf{i})) \Vert \leq e_f.
    \]
\end{ass}

Under the above assumptions, we define the following \textit{estimated} mapping that corresponds to the iterates of \ZOIO:
\begin{align}
\hat{\bm{\psi}_{p,d}}^{(k)}(\mbf{z}^{(k)}) = \begin{bmatrix}
    \partial_{\mbf{i}} \mathcal{L}^{(k)}_{p,d}(\mbf{i}^{(k)}, \bm{\lambda}^{(k)}) + O( \epsilon + \sigma^2 + e_f + e_y ) \\
    - h(g^{(k)}_{\mbf{u}^{(\star,k)}}( \mbf{i}^{(k)})) + d \bm{\lambda}
\end{bmatrix}.
\end{align}
Using \cite[Lemma 4]{Bernstein:19}, we have the following bound for $\hat{\bm{\psi}_{p,d}}^{(k)}$:

\begin{lem} \label{lem:psi-error}
For some constant $e_{\bm{\psi}} = O( \epsilon + \sigma^2 + e_f + e_y )$, 
\begin{equation}
    \left\Vert \bm{\psi}_{p,d}^{(k)}(\mbf{z}^{(k)}) - \hat{\bm{\psi}_{p,d}}^{(k)}(\mbf{z}^{(k)}) \right\Vert_2 \leq e_{\bm{\psi}} \quad \forall k \geq 1.
\end{equation}
\end{lem}

By bounding this error, we can again use \cite[Theorem 4]{Bernstein:19}, to show that the iterates of \ZOIO exhibit the following tracking performance.

\begin{theorem} \label{thm:ZOIO-convergence-timevarying}
Assume that there exists a finite scalar $\omega$ such that $\sup_{k \ge 1} \omega^{(k)} \le \omega$, and let the step size $\epsilon$ be chosen such that $0 < \epsilon < \nicefrac{2\eta_{\bm{\psi}}}{L_{\bm{\psi}}^2}$.  Then the sequence of iterates $\{ \mbf{z}^{(k)} \}$ converges Q-linearly to  $\{ \mbf{z}^{(\star, k)} \}$ up to an asymptotic error bound given by
\begin{equation}
\lim_{k \to \infty} \sup \Vert \mbf{z}^{(k)} - \mbf{z}^{(\star, k)} \Vert_2 \leq \frac{ e_{\bm{\psi}} \epsilon + \omega}{1-\sqrt{1- 2\epsilon \eta_{\bm{\psi}} + \epsilon^2 L_{\bm{\psi}}^2}},
\end{equation}
where $e_{\bm{\psi}}$ is the bound in \sref{Lemma}{lem:psi-error}.
\end{theorem}

Intuitively, compared to the bound for \FOIO, the bound in \autoref{thm:ZOIO-convergence-timevarying} adds a dependence on the aggregate error due to the zero-order gradient approximation (namely, $e_{\bm{\psi}}$).
Taken together, the bounds in \sref{Theorems}{thm:FOIO-convergence-timevarying} and \ref{thm:ZOIO-convergence-timevarying} imply that the optimization algorithms presented in \autoref{sec:opt} retain convergence guarantees in the time-varying case as long as the problem satisfies some intuitive conditions (e.g., convexity, Lipschitz continuity).  While these conditions will often not hold in empirical applications, in the next section we implement these techniques towards the motivating application of distribution grid voltage control, evaluating their empirical convergence and tracking performance when these conditions are not met.

\section{Case Study: Distribution Grid Voltage Control}
\label{sec:exp}
In this section, we evaluate the incentive mechanism and optimization algorithms from \autoref{sec:opt} on a realistic distribution feeder.  

\subsection{Setup} \label{sec:exp-setup}

We simulate the IEEE 33-bus radial distribution network~\cite{Baran:89} using the \texttt{pandapower} Python library.  In these experiments, we assume that the PQ (load) buses are exclusively consumers, so the only power generation is from the slack bus, and keeping voltage magnitudes above the lower bound is the primary safety constraint.  We also assume that only the \textit{active power} loads are controllable, i.e., that reactive loads are \textit{incentive-agnostic} -- this is motivated by practice since reactive power is typically much more difficult to control and incentivize.
We note, however, that the algorithms we present are designed to capture more general cases (e.g., considering both active and reactive power).

For the PQ buses, we randomly assign 32 loads from the Smart$^\star$ data set~\cite{Barker:12}, which includes one year's worth of 1 minute load measurements for 114 apartments.  We choose an instance (i.e., a time index in the data) such that the \textit{base loads} satisfy voltage lower and upper bounds of $\underline{\mbf{v}} = 0.9$ p.u. and $\overline{\mbf{v}} = 1.1$ p.u., respectively.  These base loads are used as the linearization point $(\mbf{p}^\star, \mbf{q}^\star)$ for the LinDistFlow model.  %
We generate an increased demand by iteratively \textit{increasing the load} at PQ buses, multiplying each load with a random scalar between $1$ and $1.1$; each iteration increases the active power demand $\overline{\mbf{p}}$ by up to 10\%, and this process repeats until more than five nodal voltage magnitudes drop below their $0.9$ p.u. lower bound.  The added demand is denoted by $\bm{\delta} \coloneqq \overline{\mbf{p}} - \mbf{p}^\star \in \mathbb{R}_+^n$.

\subsubsection{Algorithm implementations}

We implement each algorithm discussed in \autoref{sec:opt}, along with a few techniques that solve for the ``optimal baseline solution'' with full access to $g_{\mbf{u}^\star}(\mbf{i})$.

\noindent\textbf{Iterative heuristic. \ } As a heuristic comparison, we implement the iterative increase algorithm (\autoref{alg:iii}, \texttt{III}), setting $\epsilon = 0.1$.  While the algorithm does not have access to $g_{\mbf{u}^\star}(\mbf{i})$, it has access to ``accurate measurements'' of $g_{\mbf{u}^\star}(\mbf{i}')$ for a particular incentive input $\mbf{i'}$ (the same extends to \DAIO, \FOIO, and \ZOIO).

\noindent\textbf{Dual ascent. \ } When the functional form of $g_{\mbf{u}^\star}(\mbf{i})$ is known, we implement the dual ascent technique (\autoref{alg:daio}, \DAIO), with a closed form primal update given by \eqref{eq:argminCF}.  Our implementation of \DAIO uses a simple time-varying step size (for the dual variables $\bm{\lambda}$), where $\epsilon_1 = 1$, and $\epsilon_{k+1}$ is incremented by $\epsilon_{k} + \nicefrac{1}{10}$ whenever the voltages are \textit{not within bounds}.

\noindent\textbf{First-order primal-dual. \ }  Our implementation of the first-order primal-dual algorithm (\FOIO, \autoref{alg:foio}) uses different step sizes for the primal and the dual updates, respectively.  For the primal update, we set a constant step size of $\varepsilon = 0.0005$.  For the dual update, we follow the same time-varying rule used in the implementation of \DAIO (i.e., $\epsilon_1 = 1$, $\epsilon_{k+1} = \epsilon_{k} + \nicefrac{1}{10}$ whenever the voltages are \textit{not within bounds}).
\FOIO takes the gradient $\nabla_{\mbf{i}} g_{\mbf{u}^\star}(\mbf{i})$ as an input -- in our experiments, we pass either the true gradient $\nabla_{\mbf{i}} g_{\mbf{u}^\star}(\mbf{i})$, or an approximation of it.  We explicitly describe the gradient passed to \FOIO in the details of each experiment.

\noindent\textbf{Zero-order approximation. \ } 
In the regime where the gradient $\nabla_{\mbf{i}} g_{\mbf{u}^\star}(\mbf{i})$ is unknown and not estimatable, we implement (\autoref{alg:zoio}, \ZOIO).  As in \FOIO, we use different step sizes for the primal and the dual updates -- for the primal update, we fix a constant step size of $\varepsilon = 0.0001$.  For the dual update, we follow a similar time-varying rule as above, setting an initial $\epsilon_1 = 1$, but we only increment $\epsilon_{k+1} = \epsilon_{k} + \nicefrac{1}{10}$ if the voltages are not within bounds \textit{and} \ZOIO has \textit{not yet found} an incentive that satisfies the constraint.
We set the exploration parameter $\sigma$ to different values based on different $g_{\mbf{u}^\star}(\mbf{i})$ functions considered -- in general, larger values of $\sigma$ allow the algorithm to converge in ``more difficult'' problems (e.g., non-smooth $g$ functions), but result in a larger optimality gap at convergence (see \autoref{thm:ZOconvergence-nontimevary} for the theoretical bound).  We set $d$ (in the augmented Lagrangian) such that $(1-\epsilon_k d) = 0.95$, amounting to a 5\% \textit{decay} in $\bm{\lambda}$ at each step, and $p$ is set to $0$.  The random exploration vector $\bm{\zeta}^{(k)}$ is resampled at each iteration, and each term is drawn uniformly from $[-1,1]$.

\noindent\textbf{Solving for the optimal baseline. \ }
We use two techniques to solve for the optimal solution (or approximate optimal solution) depending on the structure of $g_{\mbf{u}^\star}(\mbf{i})$.  Since these solvers require knowledge of $g_{\mbf{u}^\star}(\mbf{i})$ to work,  they serve as a baseline to evaluate the performance of our main feedback-based algorithms.
When $g_{\mbf{u}^\star}(\mbf{i})$ is \textit{convex}, we use CVXPY~\cite{CVXPY} to solve for the optimal incentive.  When $g_{\mbf{u}^\star}(\mbf{i})$ is \textit{linear} (i.e., $g = \overline{g}$), we reformulate \eqref{eq:sci} as an LP and use SciPy's LP solver~\cite{SciPy}.  
When $g_{\mbf{u}^\star}(\mbf{i})$ is \textit{not convex or linear}, we use the $\bm{\delta}$ and $\mbf{t}$ values from the true function to construct a linear approximation $\overline{g}$ according to \sref{Def.}{dfn:linear-g}, and solve for the \textit{approximate optimal solution} under the corresponding LP -- we find that this approximate optimal solution serves as a useful baseline.

\subsubsection{Defining incentive response $g_{\mbf{u}^\star}(\mbf{i})$} \label{sec:implementing-g}
In the main body, we highlight two experiment settings -- one where $g_{\mbf{u}^\star}(\mbf{i})$ satisfies the theoretical assumptions for convergence, and one where $g_{\mbf{u}^\star}(\mbf{i})$ adopts a more realistic form that would be expected in a real deployment.  In \autoref{apx:supp-exp}, we present results on additional types of $g_{\mbf{u}^\star}(\mbf{i})$ beyond the two settings presented here.

In every case, we start with the \textit{increased demand} $\bm{\delta}$ for the underlying instance -- this fixes a point for $g$ when the incentive is $\mbf{0}$, namely $g_{\mbf{u}^\star}(\mbf{0}) = \mbf{u}^\star + \bm{\delta}$.  Recall that $\mbf{u}^\star$ corresponds to a setpoint for the active power demand (i.e., $\mbf{p}^\star$).  We then generate a random threshold vector $\mbf{t}$, where each element is drawn from a uniform distribution on $[0,1]$.  From \sref{Assumption}{asm:threshold}, this fixes a second point for $g$ when the incentive is $\mbf{t}$, namely $g_{\mbf{u}^\star}(\mbf{t}) = \mbf{u}^\star$.  Between these two points, we consider the following functional forms:

\noindent\textbf{Quadratic-convex case. \ } 
In this case, we implement $g_{\mbf{u}^\star}(\mbf{i})$ to adopt the \textit{quadratic-convex} form considered in \autoref{sec:quad-convex-CF-DAIO}.  Recall that we have $g_{\mbf{u}^\star}(\mbf{i}) = \mbf{u}^\star + \mbf{b} \odot (\mbf{i} - \mbf{t})^{\circ 2}$ -- to satisfy the boundary conditions at $\mbf{i} = \mbf{0}$ and $\mbf{i} = \mbf{t}$, we set $\mbf{b} = \bm{\delta} \oslash \left( \mbf{t}^{\circ 2} \right) $.  Note that this implementation of $g$ satisfies the necessary assumptions for convergence in \autoref{sec:opt}. %

\begin{wrapfigure}{r}{0.5\textwidth}
    \vspace{-1em}
  \includegraphics[width=\linewidth]{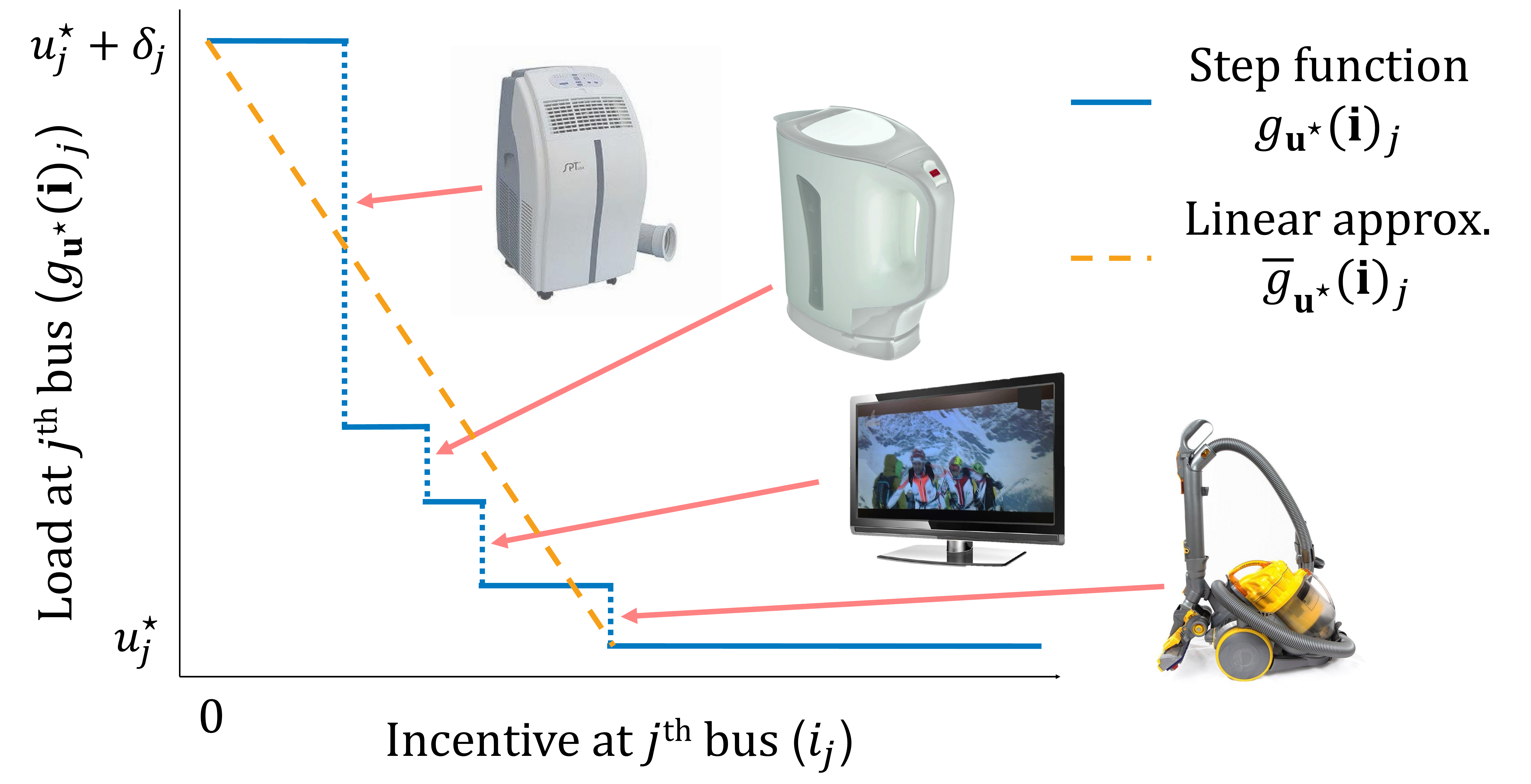} \vspace{-2em}
    \caption{An annotated example of a step $g$ function at a single bus $j$.  Each step corresponds to a ``controllable device''.}
    \label{fig:step} \vspace{-2em}
\end{wrapfigure}

\noindent\textbf{Step function case. \ }
In this more realistic case, we construct a $g_{\mbf{u}^\star}(\mbf{i})$ to model \textit{discrete controllable devices} -- in the context of load shedding for voltage control, these can be understood as devices that an incentivized end-user is willing to turn off or allow the SO to control.
To model this case, $g_{\mbf{u}^\star}(\mbf{i})$ is implemented as a collection of step functions, one for each PQ bus (i.e., end-user).  In \autoref{fig:step}, we plot an example of such a step function for a single bus.  At a high-level, as the incentive to a given bus is increased, the demand from that bus decreases in a non-smooth fashion as devices are individually ``turned off''.

For the $j^{\text{th}}$ bus, we specify a number of ``devices'' $D \in \mathbb{N}$ at initialization.  We start with a single ``step'' that stays at $u^\star_j + \delta_j$ until the incentive is increased to $t_j$, at which point it drops to $u^\star_j$.  We then iteratively add new ``steps'' in between existing ones using randomization -- e.g., for the first iteration, we choose an incentive breakpoint $i_j'$ from a uniform distribution on $[0,t_j]$, and a demand breakpoint $u'_j$ from a uniform distribution on $[u^\star_j, u^\star_j + \delta_j]$.  After this iteration, $g$ starts at $u^\star_j + \delta_j$ until the incentive is $\geq$ $i_j'$, at which point it drops to $u'_j$ and stays there until the incentive reaches $t_j$.  This iterative process repeats until the number of changes in the value of $g_{\mbf{u}^\star}(\mbf{i})_j$ is equal to $D$, the desired number of devices.
As a limiting case of this step function idea (i.e., when $D \to \infty$), in these experiments we also consider the case where $g_{\mbf{u}^\star}(\mbf{i})$ is linear (i.e., $g_{\mbf{u}^\star}(\mbf{i}) = \overline{g}_{\mbf{u}^\star}( \mbf{i})$, see \sref{Def.}{dfn:linear-g}). 

\subsubsection{Time-varying case: defining $g^{(k)}_{\mbf{u}^{(\star,k)}}( \mbf{i})$} \label{sec:timeVaryingGDefExp}

Drawing on results in \autoref{sec:timevarying}, we additionally consider the case where the underlying $g_{\mbf{u}^\star}(\mbf{i})$ \textit{changes over time}.  To implement a ``realistic'' time-varying instance, we start with the step $g$ function defined in the previous section -- specifically, we let $g^{(0)}_{\mbf{u}^{(\star,0)}}( \mbf{i})$ be a collection of step functions with $D$ devices at each bus, as outlined above.  

In contrast to the ``fast time-scale'' of individual iterations in \FOIO and \ZOIO, we consider a ``slow time-scale'' for the time-varying $g$ function, where changes happen at e.g., a minute-level granularity.  In what follows, we typically assume that 1,000 ``fast'' iterations correspond to $1$ minute, and that $g_{\mbf{u}^\star}(\mbf{i})$ changes once per minute.

We construct time-varying instances that span $100$ minutes.  We model ``events'' at each bus (e.g., devices arriving or leaving) as a Poisson birth-death process where events happen once every $2.5$ minutes on average. For each event, we choose whether to add or remove with equal probability -- the only exception is if a particular bus has only 2 controllable devices remaining, in which case the event is discarded.
When a device is \textit{added}, we randomly generate a tuple representing the increase in demand and the increase in the required incentive.  The ranges for each of these increases are determined by the devices present at time step $k=0$.  The corresponding device is added ``from the left'' by adding a step at the beginning of the function -- this results in a new peak demand when the incentive is $0$ (i.e., a new $\bm{\delta}$) and the incentives for other devices are incremented accordingly (i.e., $\mbf{t}$ changes) -- we henceforth index $\bm{\delta}^{(k)}$ and $\mbf{t}^{(k)}$ to capture this.
When a device is \textit{removed}, the first device (i.e., ``from the left'') is removed, so the new peak demand $\bm{\delta}$ is given by the second step from the left, and the incentives for other devices are decremented accordingly.

\noindent\textbf{Quadratic-convex time-varying case. \ }
Motivated by the theoretical results, namely \autoref{thm:FOIO-convergence-timevarying} and \ref{thm:ZOIO-convergence-timevarying}, we additionally consider the case where $g_{\mbf{u}^\star}(\mbf{i})$ changes over time while satisfying the necessary assumptions for convergence.  For the sake of continuity, we implement time-varying \textit{quadratic-convex} $g$ functions that are based on the device arrival/departure model described above.
Concretely, we let $g^{(k)}_{\mbf{u}^{(\star,k)}}( \mbf{i}) = \mbf{u}^\star + \mbf{b}^{(k)} \odot (\mbf{i} - \mbf{t}^{(k)})^{\circ 2}$, where $\mbf{b}^{(k)} = \bm{\delta}^{(k)} \oslash (\mbf{t}^{(k) \circ 2})$, and the time-varying quantities $\bm{\delta}^{(k)}$ and $\mbf{t}^{(k)}$ are inherited from a sequence of time-varying step functions as defined above.

\begin{figure*}[t]
    \vspace{-0.5em}
    \minipage{0.48\textwidth}
    \includegraphics[width=\linewidth]{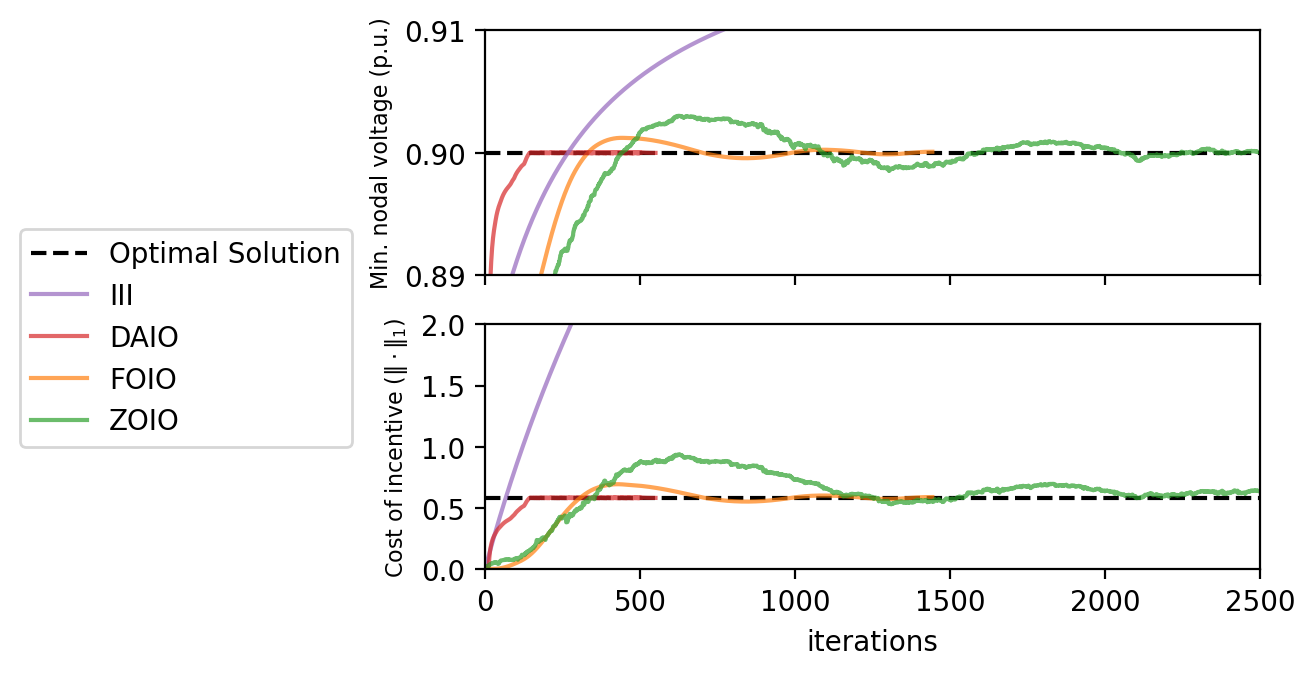}\vspace{-1em}
    \caption{ Min. nodal voltage magnitude (top) and incentive cost (bottom) vs. iterations for \textit{quadratic-convex} stationary $g$ function experiment. }\label{fig:quad-convex}
    \endminipage\hfill
    \minipage{0.48\textwidth}
    \centering
    \includegraphics[width=\linewidth]{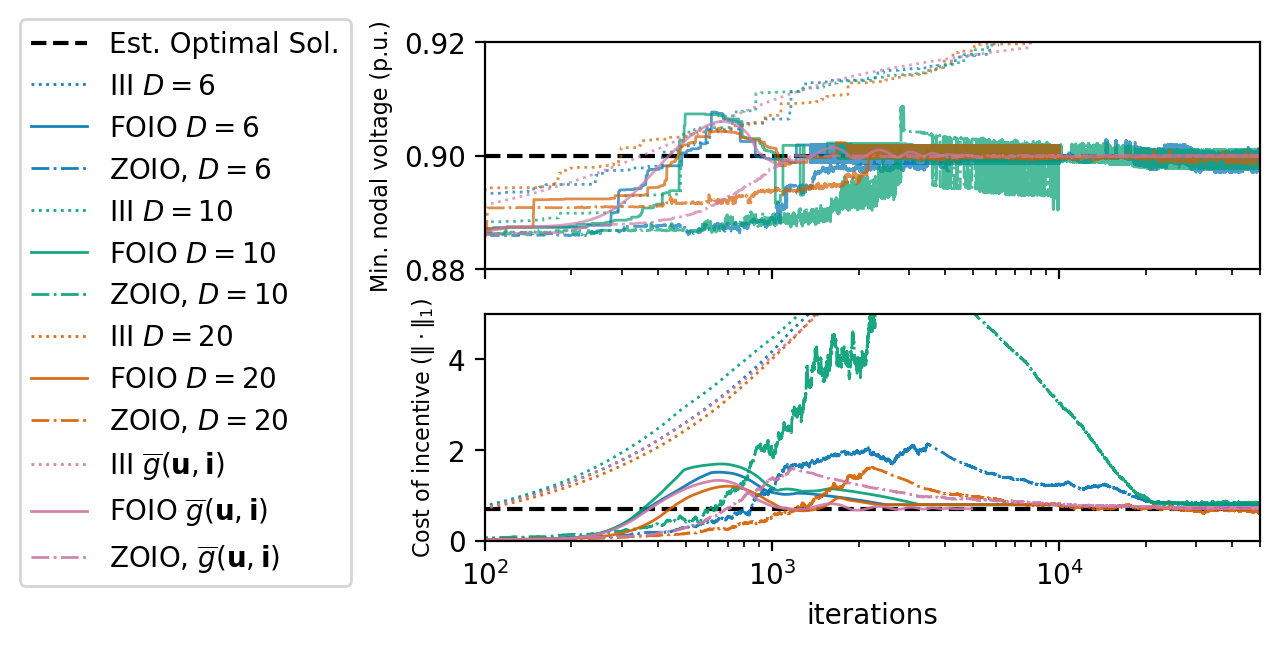}\vspace{-1em}
    \caption{ Min. nodal voltage magnitude (top) and incentive cost (bottom) vs. iterations for stationary $g$ step function experiment ($D$ indicates \# of controllable devices at each bus). }\label{fig:step-linear}
    \endminipage
    \vspace{-1em}
\end{figure*}

\subsection{Stationary Results} \label{sec:exp-results-station}

In this section we present results for experiments where the $g_{\mbf{u}^\star}(\mbf{i})$ function is \textit{stationary}.  We compare the algorithms we test by comparing the minimum nodal voltage magnitude (p.u.) and cost of the incentive at each iteration.  In each figure, we also plot the true optimal solution or an approximate optimal solution as a benchmark.

In \autoref{fig:quad-convex}, we present results of an experiment with \textit{quadratic-convex} $g$ functions.  In this experiment, $\FOIO$ is given the true gradient $\nabla_{\mbf{i}} g_{\mbf{u}^\star}(\mbf{i})$, and $\ZOIO$'s exploration parameter is set to $\sigma = 0.005$.  Since the $g$ functions in this experiment satisfy the closed-form dual ascent update in \eqref{eq:argminCF}, we include \DAIO in this experiment as a baseline. 
The results of this experiment largely mirror the behavior predicted by the theoretical results.  Given substantial information about $g_{\mbf{u}^\star}(\mbf{i})$, \DAIO converges to the optimal solution within a few hundred iterations.  When given accurate gradient information about $g$, \FOIO is able to converge to the optimal within 1,500 iterations.  \ZOIO's model-free nature slows its convergence substantially since it has to estimate gradients at each step -- still, within a few thousand iterations, \ZOIO's iterates stabilize within a small region around the optimal solution, as predicted by \autoref{thm:ZOconvergence-nontimevary}.  Interestingly, our heuristic iterative increase method (\texttt{III}) is second only to \DAIO in terms of the time to find a \textit{feasible point} for voltage magnitudes -- however, the corresponding incentive found by \texttt{III} is quite suboptimal, this is the case because \texttt{III} cannot consider inter-bus relationships between voltages in the network.

\begin{figure*}[t]
    \minipage{\textwidth}
    \centering
    \includegraphics[width=0.35\linewidth]{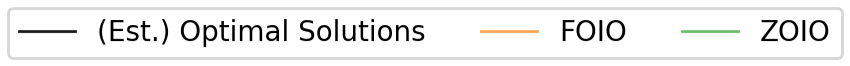}
    \endminipage\hfill
    \minipage{0.48\textwidth}
    \centering
    \includegraphics[width=0.85\linewidth]{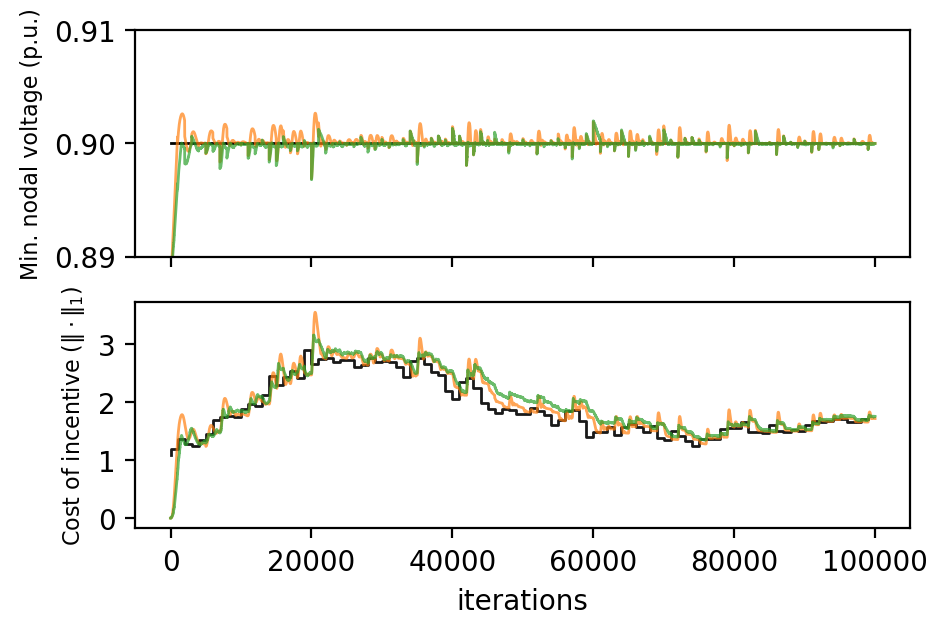}\vspace{-1em}
    \caption{ Min. nodal voltage magnitude (top) and incentive cost (bottom) vs. iterations for \textit{quadratic-convex} time-varying $g$ function experiment. }\label{fig:quad-convex-TV}
    \endminipage\hfill
    \minipage{0.48\textwidth}
    \centering
    \includegraphics[width=0.85\linewidth]{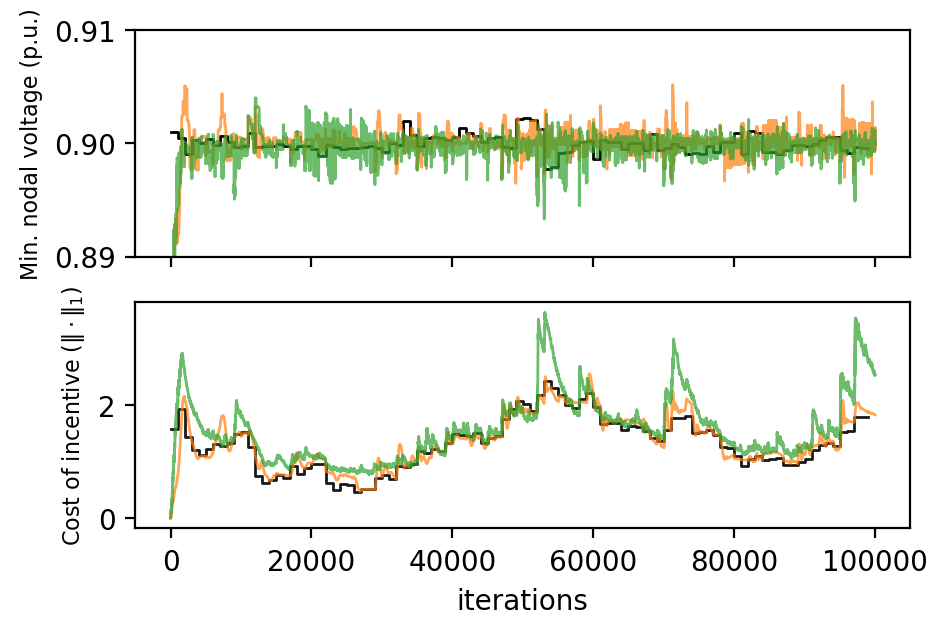}\vspace{-1em}
    \caption{ Min. nodal voltage magnitude (top) and incentive cost (bottom) vs. iterations for time-varying $g$ step function experiment ($D = 6$ controllable devices on average). }\label{fig:step-linear-TV}
    \endminipage\hfill
\end{figure*}

For the more realistic setting, in \autoref{fig:step-linear} we present results for $g$ functions that are a collection of step functions as outlined in \autoref{sec:implementing-g}, with varying numbers for $D$ (the number of controllable devices at each bus).  In this experiment, $\FOIO$ is given $\nabla \overline{g}_{\mbf{u}^\star}(\mbf{i})$ (i.e., the gradient of the linear approximation $\overline{g}$), and $\ZOIO$'s exploration parameter is set to $\sigma = 0.1$.  
In this experiment, the number of iterations required to converge on these ``more difficult'' $g$ functions increases (note the $\log$ scale on the $x$-axis). For \ZOIO, the non-smoothness of the step functions at each bus poses a challenge for two-function evaluation gradient estimation.
Surprisingly, we note that although \FOIO is given a very rough ``gradient'' approximation owing to the non-differentiability of the step $g$ function, it is able to converge to a near-optimum relatively quickly.

Motivated by $\FOIO$'s performance in the previous experiment and the theoretical result in \autoref{thm:FOIO-convergence-bad-T}, we further investigate its convergence performance when it is given ``incorrect'' gradients.  Recall that $\overline{g}_{\mbf{u}^\star}( \mbf{i})$ is fully parameterized by $\bm{\delta}$ and $\mbf{t}$ (see \sref{Def.}{dfn:linear-g}).  In a real deployment, $\bm{\delta}$ could be measured, but $\mbf{t}$ is private information (i.e.,  unknown to the SO).  Thus, we consider the case where $\FOIO$ is given estimated gradients of $\overline{g}$ based on a very coarse estimate of $\mbf{t}$ from e.g., prior data.  Specifically, we construct a gradient $\widehat{\nabla} \overline{g'}_{\mbf{u}^\star}(\mbf{i})$ according to an  estimate $\mbf{t'}(\cdot) \in \mathbb{R}^n$, where $\mbf{t'}( x ) \coloneqq \mbf{1} \cdot x$.  Plainly, $\mbf{t'}(x)$ is an estimated threshold vector that assumes end-users will meet the desired setpoint $\mbf{u}^\star$ when they are each given incentive $x \in \mathbb{R}_+$, where $x$ is e.g., a coarse estimate based on historical data.

\begin{wrapfigure}{r}{0.5\textwidth}
    \vspace{-1em}
  \includegraphics[width=\linewidth]{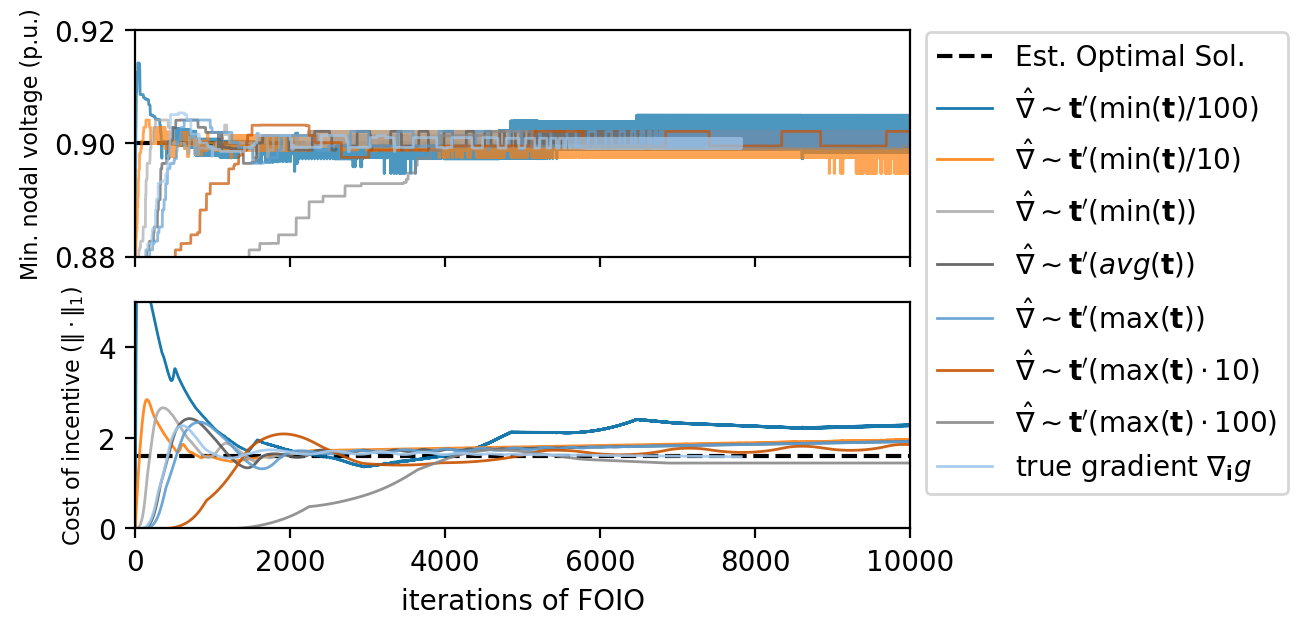}\vspace{-1em}
     \caption{ Min. nodal voltage magnitude (top) and total cost of incentive (bottom) vs. iterations of \FOIO on a step $g$ function instance with $D = 6$ controllable devices at each bus.  Each instance of \FOIO is given an \textit{incorrect gradient} $\widehat{\nabla} \overline{g'}_{\mbf{u}^\star}(\mbf{i})$, where $\overline{g'}$ is constructed according to a threshold vector estimate $\mbf{t'}(\cdot)$. }\label{fig:grad-est} \vspace{-1em}
\end{wrapfigure}

In \autoref{fig:grad-est}, we find that $\FOIO$ converges to a near-optimal incentive even when it is given gradients based on poor estimates of the threshold vector $\mathbf{t}$.  For instance, $\mbf{t'}(\min(\mathbf{t})/100)$ assumes the buses are willing to match the SO's setpoint $\mbf{u}^\star$ when given an individual incentive that is $\nicefrac{1}{100}^\text{th}$ as large as the smallest element in the true threshold vector $\mbf{t}$.  Even in this case, on a step $g$ function with $D = 6$ devices at each bus, \FOIO's incentive at convergence is just 22.1\% costlier than that of the estimated optimal solution.
These experimental results, combined with \autoref{thm:FOIO-convergence-bad-T}, suggest that \FOIO is robust to gradient error under the structure of Problem \eqref{eq:sci}, successfully converging to a near-optimum when given any estimate that roughly aligns with the true shape of $g_{\mbf{u}^\star}(\mbf{i})$.  Under e.g., monotonicity assumptions (see \sref{Assumption}{asm:monotone}), such \textit{coarse approximations} may be relatively straightforward to obtain in practice.  %

\subsection{Time-Varying Results}

In this section we present results for experiments where $g_{\mbf{u}^\star}(\mbf{i})$ is \textit{time-varying} according to the implementation described in \autoref{sec:timeVaryingGDefExp}.  As in \autoref{sec:exp-results-station}, we start by showing results for a setting that matches the theoretical conditions for convergence before moving to the more realistic ``step function'' case.  We compare the algorithms we test by plotting the minimum nodal voltage magnitude (p.u.) and the cost of the incentive over a fixed number of iterations (100,000) that corresponds to 100 minutes of the ``slow time-scale'' $g$ function evolution process.

In \autoref{fig:quad-convex-TV}, we present results of an experiment with \textit{quadratic-convex} $g$ functions.  In this experiment, $\FOIO$ is given the true gradient $\nabla_{\mbf{i}} g_{\mbf{u}^\star}(\mbf{i})$, and $\ZOIO$'s exploration parameter is set to $\sigma = 0.005$.  We only show results for \FOIO and \ZOIO in this plot to aid in readability, although an expanded set of time-varying experiments in \autoref{apx:supp-exp} includes results for \texttt{III} and \DAIO.
As we observed in the stationary case, both \FOIO and \ZOIO are adept at tracking the time-varying optimal solution when the underlying $g$ functions are quadratic-convex.  Note that the optimal solution changes every 1,000 iterations -- these transitions can be observed in the voltage graph, as the time-varying setting sometimes causes a previously feasible incentive to no longer be feasible.

In the more realistic setting where discrete devices arrive and depart over time, in \autoref{fig:step-linear-TV} we present results for $g$ step functions implemented according to \autoref{sec:timeVaryingGDefExp} -- in this experiment, each bus has an \textit{average} of $6$ controllable devices, $\FOIO$ is given $\nabla \overline{g}_{\mbf{u}^\star}(\mbf{i})$ (i.e., the gradient of the linear approximation $\overline{g}$), and $\ZOIO$'s exploration parameter is set to $\sigma = 0.1$. 
Due to the challenges of the time-varying step function, both \FOIO and \ZOIO visibly run into more difficulty tracking the optimal solution in this experiment.  Despite this though, \FOIO and \ZOIO's average incentives over time are only 2.65\% and 19.1\% more costly compared to the estimated optimal solution, respectively.  Similar to the stationary case, since \ZOIO must estimate the gradient at each iteration, it is more susceptible to iterates that go in the ``wrong direction'' -- in the graph, these are visible as large ``spikes'' in the incentive cost that require some iterations to recover from.

\section{Conclusion}
\label{sec:conclusion}
We study an individualized incentive problem motivated by the problem of behind-the-meter control of emerging assets such as DERs in electric grids.  Our formulation provides a layer of abstraction over both the details of the incentive and the environment's response (e.g., how users respond to a given incentive), distinguishing it from prior work.  For the realistic case in which a system operator does not have information about e.g., the effect of a certain incentive, we propose feedback-based optimization algorithms (\FOIO and \ZOIO) that iteratively update incentives while ensuring system constraints are met.  We prove theoretical bounds on the convergence of these techniques under some necessary assumptions on the problem (e.g., convexity, Lipschitzness).  Furthermore, in our case study, we show that the proposed algorithms remain effective when the environment's response does not satisfy these necessary assumptions, modeling a more realistic scenario.

A number of questions remain for future work.  While our case study instantiates the incentive problem for a stylized example of voltage control in distribution grids, it would be worthwhile to consider a more complicated system with more constraints, such as net power injection limits.  In a similar vein, our study assumed that the system operator has knowledge of system information such as e.g., the grid's topology -- in reality, situations may arise where this is not the case, motivating further inquiry into what can be accomplished when both the environment's response and the details of the system must be ``learned'' in a feedback-based manner.

\section*{Acknowledgments \& Disclaimers}
This work was authored in part by the National Renewable Energy Laboratory (NREL), operated by Alliance for Sustainable Energy, LLC, for the U.S. Department of Energy (DOE) under Contract No. DE-AC36-08GO28308. This work was supported by the Laboratory Directed Research and Development (LDRD) Program at NREL. The views expressed in the article do not necessarily represent the views of the DOE or the U.S. Government. The U.S. Government retains and the publisher, by accepting the article for publication, acknowledges that the U.S. Government retains a nonexclusive, paid-up, irrevocable, worldwide license to publish or reproduce the published form of this work, or allow others to do so, for U.S. Government purposes.

This material is based upon work supported by the U.S. Department of Energy, Office of Science, Office of Advanced Scientific Computing Research, Department of Energy Computational Science Graduate Fellowship under Award Number DE-SC0024386.

This report was prepared as an account of work sponsored by an agency of the United States Government. Neither the United States Government nor any agency thereof, nor any of their employees, makes any warranty, express or implied, or assumes any legal liability or responsibility for the accuracy, completeness, or usefulness of any information, apparatus, product, or process disclosed, or represents that its use would not infringe privately owned rights. Reference herein to any specific commercial product, process, or service by trade name, trademark, manufacturer, or otherwise does not necessarily constitute or imply its endorsement, recommendation, or favoring by the United States Government or any agency thereof. The views and opinions of authors expressed herein do not necessarily state or reflect those of the United States Government or any agency thereof.

\printbibliography

\appendix
\section*{Appendix}

\section{Supplemental Experiments} \label{apx:supp-exp}

In this section, we present results for a few experiments that are not included in the main body.  While much of the setup is inherited from \autoref{sec:exp-setup}, we start by detailing the setup of additional settings considered here. 

\subsection{Supplemental Setup}

\subsubsection{Additional incentive responses $g_{\mbf{u}^\star}(\mbf{i})$} \label{apx:more-g}

In these appendix experiments, we consider an expanded set of smooth $g$ functions that are detailed below.

\paragraph{Polynomial-convex case.} 
In this case, we implement $g_{\mbf{u}^\star}(\mbf{i})$ to adopt a polynomial convex form that generalizes the quadratic-convex case already considered in \autoref{sec:quad-convex-CF-DAIO}.  We set $g_{\mbf{u}^\star}(\mbf{i}) = \mbf{u}^\star + \mbf{b} \odot (\mbf{i} - \mbf{t})^{\circ y}$ for some arbitrary exponent $y \in \{2, 4, 6, 8, \dots \}$, where $\bm{\delta}$ and $\mbf{t}$ are constant across all $y$.  To satisfy the boundary conditions at $\mbf{i} = \mbf{0}$ and $\mbf{i} = \mbf{t}$, we set $\mbf{b} = \bm{\delta} \oslash \left( \mbf{t}^{\circ y} \right) $.  

\paragraph{Polynomial-concave case.}
In this case, we implement $g_{\mbf{u}^\star}(\mbf{i})$ to adopt a polynomial \textit{concave} form that is not explicitly considered in the main body.  Concretely, we set $g_{\mbf{u}^\star}(\mbf{i}) = \mbf{u}^\star + \bm{\delta} - \mbf{b} \odot (\mbf{i})^{\circ y}$ for an even exponent $y \in \{2, 4, 6, 8, \dots \}$, where $\bm{\delta}$ and $\mbf{t}$ are constant across all $y$. To satisfy the boundary conditions at $\mbf{i} = \mbf{0}$ and $\mbf{i} = \mbf{t}$, we set $\mbf{b} = \bm{\delta} \oslash \left( \mbf{t}^{\circ y} \right) $ -- note that in this case, because $g$ is concave, we have that \eqref{eq:sci} is \textit{non-convex}.

\subsubsection{Time-varying $g^{(k)}_{\mbf{u}^{(\star,k)}}( \mbf{i})$ from data} \label{apx:more-timeVaryingGDefExp}

Supplementing the device arrival/departure model studied in the main body (see \autoref{sec:exp-setup}), in these appendix experiments we also consider a time-varying $g^{(k)}_{\mbf{u}^{(\star,k)}}( \mbf{i})$ that is defined based on data.  Recall that the Smart$^\star$ data set~\cite{Barker:12} includes a time series of 1 minute load measurements for 114 real apartments over a period of one year.  To define a time-varying $g^{(k)}$ based on this data, we start by defining an initial \textit{quadratic-convex} $g^{(0)}$ exactly according to the existing technique in \autoref{sec:exp-setup}.  In doing so, we note the \textit{time index} of the chosen load profile in the data (where there are 525,600 such ``time slots'' over the course of one year of data).

At regular intervals (e.g., every 100 iterations), $g^{(k)}_{\mbf{u}^{(\star,k)}}( \mbf{i})$ is updated  as follows.  We first retrieve new loads for each PQ bus from the next ``minute'' of the underlying data.  To ``smooth out'' large fluctuations, we set a combination parameter $\alpha$ (typically $\alpha = 0.5$) and create a new ``base load'' for all buses by taking $(1-\alpha)$ of the existing load and $\alpha$ from the new load.  Finally, we generate a new ``increased demand'' $\bm{\delta}^{(k)}$ by following the iterative random process described in \autoref{sec:exp-setup}.  Each time $g^{(k)}_{\mbf{u}^{(\star,k)}}( \mbf{i})$ is updated, a new set point $\mbf{u}^{(\star, k)}$ is generated by using the new ``base loads'' to refresh the linearization point $(\mbf{p}^\star, \mbf{q}^\star)$.

\subsection{Supplemental Stationary Results}

\begin{figure*}[h]
    \vspace{-0.5em}
    \minipage{0.48\textwidth}
    \includegraphics[width=\linewidth]{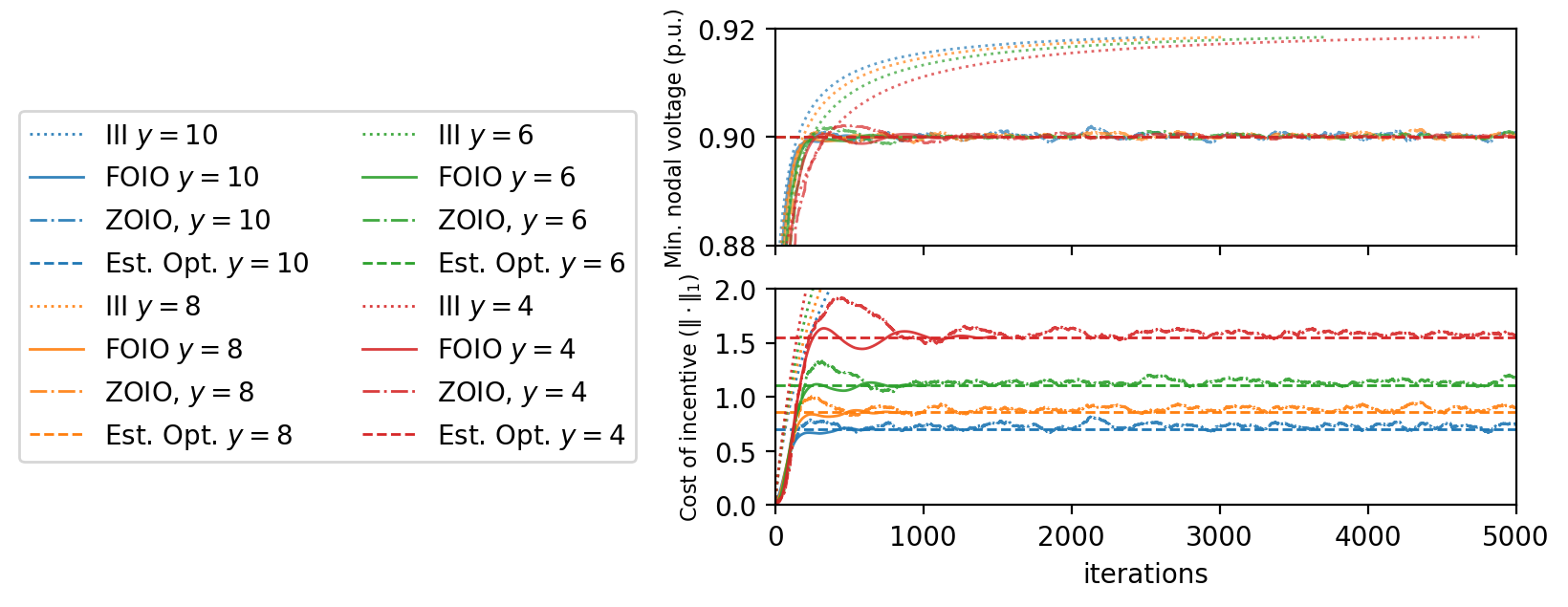}\vspace{-1em}
    \caption{ Min. nodal voltage magnitude (top) and incentive cost (bottom) vs. iterations for \textit{polynomial-convex} stationary $g$ function experiment. }\label{fig:poly-convex}
    \endminipage\hfill
    \minipage{0.48\textwidth}
    \centering
    \includegraphics[width=\linewidth]{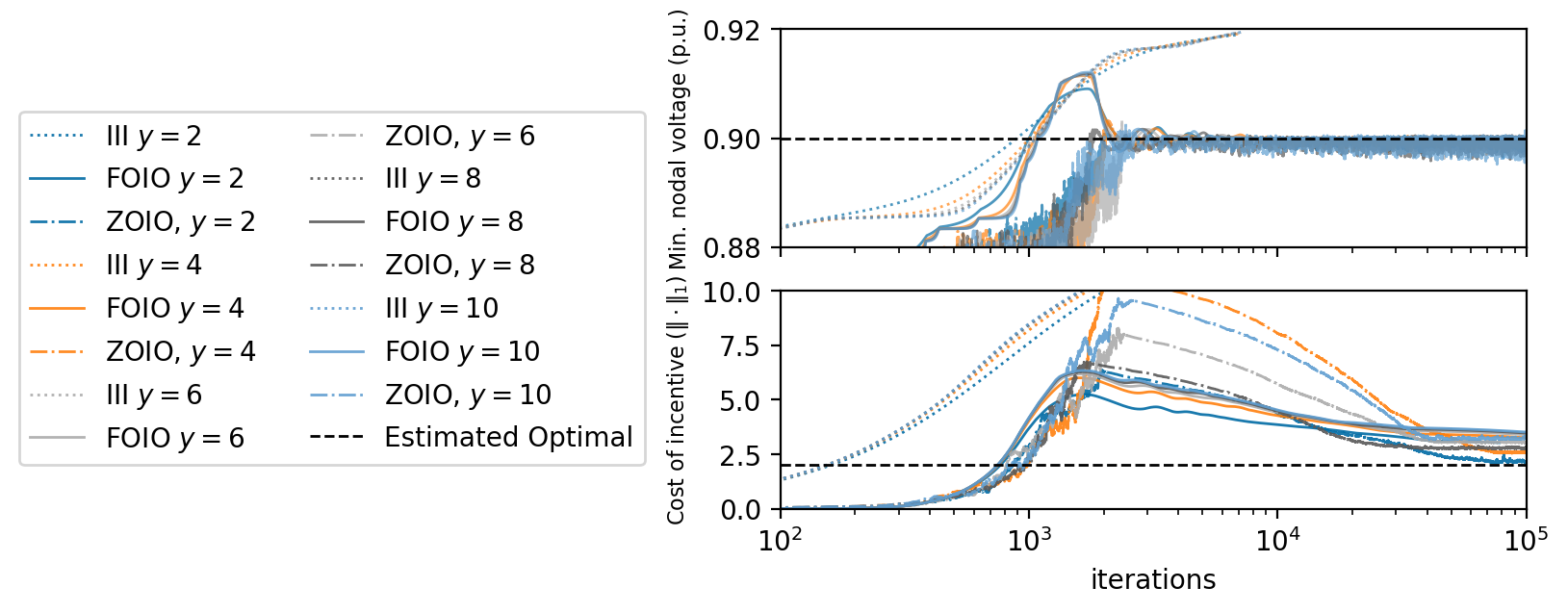}\vspace{-1em}
    \caption{ Min. nodal voltage magnitude (top) and incentive cost (bottom) vs. iterations for \textit{polynomial-concave} stationary $g$ function experiment. }\label{fig:poly-concave}
    \endminipage
    \vspace{-1em}
\end{figure*}

In \autoref{fig:poly-convex}, we present results of an experiment with \textit{polynomial-convex} $g$ functions.  We set $y = \{10, 8, 6, 4\}$, give $\FOIO$ the true gradient $\nabla_{\mbf{i}} g_{\mbf{u}^\star}(\mbf{i})$, and $\ZOIO$'s exploration parameter is set to $\sigma = 0.005$.  
The results of this experiment largely mirror the behavior observed in \autoref{fig:quad-convex}. Since this form of $g_{\mbf{u}^\star}(\mbf{i})$ satisfies conditions required for the theoretical results, both \FOIO and \ZOIO converge to the (near-)optimal solution.  In this case, since each $g_{\mbf{u}^\star}(\mbf{i})$ is convex, we use CVXPY to solve for the optimal solution -- note that for different values of $y$, the optimal incentive changes according to the convexity of the underlying $g$ function.

In \autoref{fig:poly-concave}, we present results of an experiment with \textit{polynomial-concave} $g$ functions.  We set $y = \{2, 4, 6, 8, 10\}$, give $\FOIO$ an estimated gradient according to the linear approximation $\nabla_{\mbf{i}} \overline{g}_{\mbf{u}^\star}( \mbf{i})$, and $\ZOIO$'s exploration parameter is set to $\sigma = 0.1$.  Since this form of $g_{\mbf{u}^\star}(\mbf{i})$ causes \eqref{eq:sci} to be non-convex, the results of this experiment mirror the behavior observed for $g$ step functions observed in \autoref{fig:step-linear}.  We note that the estimated optimal solution is based on the linear approximation   $\overline{g}_{\mbf{u}^\star}( \mbf{i})$, and thus constitutes a \textit{lower bound} on the true optimal incentive cost by the structure of the problem.

\subsection{Supplemental Time-Varying Results}

\begin{figure*}[h]
    \minipage{0.48\textwidth}
    \centering
    \includegraphics[width=0.9\linewidth]{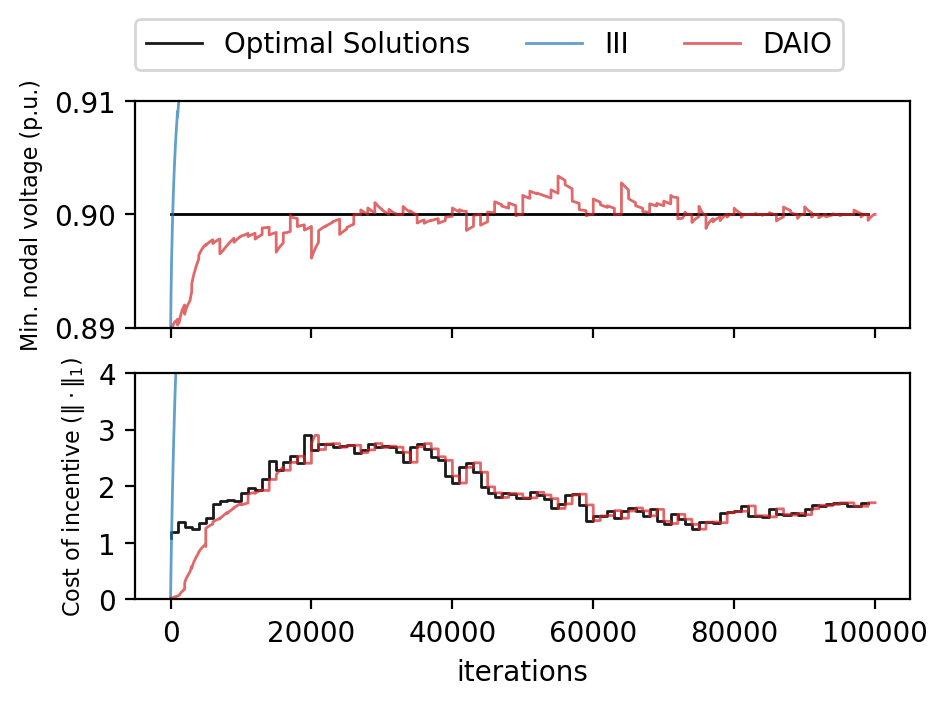}\vspace{-1em}
    \caption{ Min. nodal voltage magnitude (top) and incentive cost (bottom) vs. iterations for \texttt{III} and \DAIO algorithms on \textit{quadratic-convex} time-varying $g$ function experiment. }\label{fig:quad-convex-TV-III}
    \endminipage\hfill
    \minipage{0.48\textwidth}
    \centering
    \includegraphics[width=0.95\linewidth]{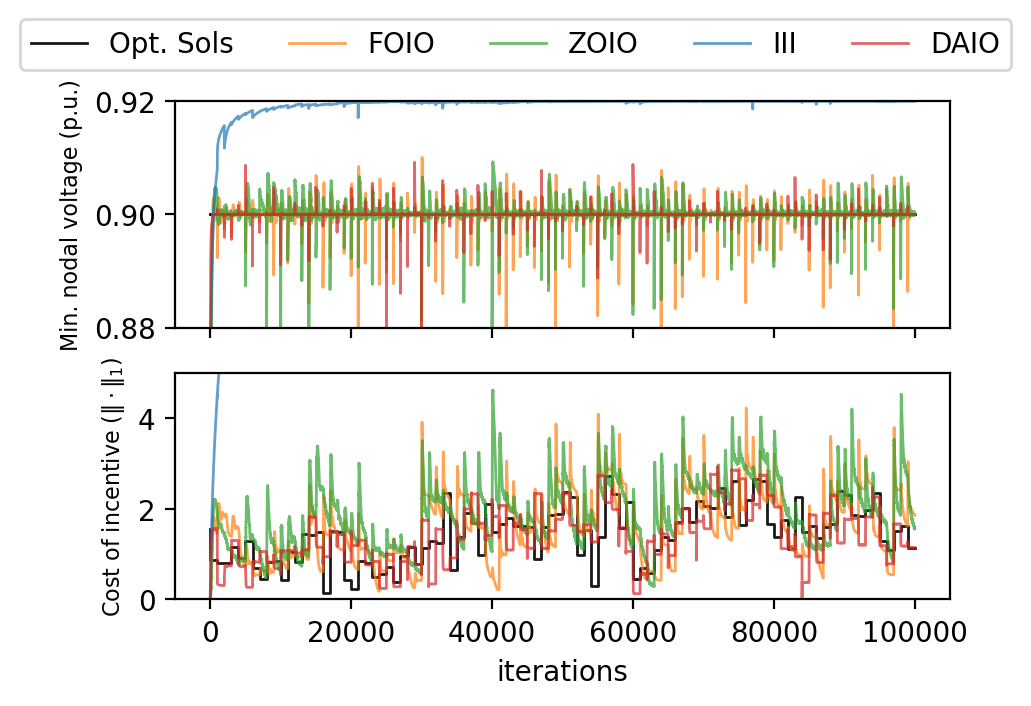}\vspace{-1em}
    \caption{ Min. nodal voltage magnitude (top) and incentive cost (bottom) vs. iterations for quadratic-convex $g$ functions from time-series data. }\label{fig:TV-from-data}
    \endminipage\hfill
\end{figure*}

In \autoref{fig:quad-convex-TV-III}, we present extended results for the experiment with time-varying \textit{quadratic-convex} $g$ functions presented in \autoref{fig:quad-convex-TV} in the main body.  In this figure, we show the performance of \texttt{III} and \DAIO in this time-varying regime --  unsurprisingly, \texttt{III} does relatively poorly in terms of objective value, as observed in the stationary experiments -- it keeps the voltage magnitudes well above the 0.9 p.u. lower bound, but incurs suboptimal cost to do so.  In comparison, \DAIO does relatively well tracking the optimal solution in this regime where the underlying $g$ function satisfies the conditions to use the closed-form update in \eqref{eq:closedformPrimalUpdateDAIO}.

Similarly, in \autoref{fig:TV-from-data}, we plot results for an experiment with time-varying $g^{(k)}_{\mbf{u}^{(\star,k)}}(\mbf{i})$ functions that are based on the time-series of the data set.  We find that the time-varying instances defined according to \autoref{apx:more-timeVaryingGDefExp} seem to be ``more difficult'' to track, in the sense that the optimal solution fluctuates more from step to step.  Although the ``base loads'' follow a realistic time-series evolution, the increased demand $\bm{\delta}^{(k)}$ is resampled from scratch each time that $g^{(k)}_{\mbf{u}^{(\star,k)}}( \mbf{i})$ is updated -- thus, these random fluctuations likely contribute to the volatility of the optimal solution.  Despite this, \DAIO, \FOIO, and \ZOIO are able to track with the optimal solution generally within a few hundred iterations -- intuitively, due to the larger fluctuations, both \FOIO and \ZOIO are more subject to ``spikes'' in incentive due to e.g., sudden voltage violations from the randomly changing $\bm{\delta}^{(k)}$.

\section{Deferred Proofs from \autoref{sec:opt} (Feedback Optimization Algorithms)}

In this section, we provide full proofs for the convergence and stability results stated for \DAIO, \FOIO,  and \ZOIO in \autoref{sec:opt}. 

\subsection{Proof of \autoref{thm:DAIOconvergence}} \label{apx:DAIOconvergence}

In the following, we prove \autoref{thm:DAIOconvergence}, which states that \DAIO (\autoref{alg:daio}) converges to the unique minimizer of \eqref{eq:sci} when
\begin{equation}
    0 < \epsilon < \frac{2 m(\bm{\lambda}^{(0)})}{\Vert h(g_{\mbf{u}^\star}(\mbf{i}^{(0)})) \Vert_2^2}, 
\end{equation}
where $m$ is the dual problem of \eqref{eq:sci}, i.e., $m(\bm{\lambda}) = \min_{\mbf{i}} \mathcal{L}(\mbf{i}, \bm{\lambda})$.

\begin{proof}[Proof of \autoref{thm:DAIOconvergence}]
\noindent Recall that the Lagrangian $\mathcal{L}$ and its corresponding dual function $m$ are given by 
\begin{align*}
\mathcal{L}(\mbf{i}, \bm{\lambda}) &= c(\mbf{i}) + \bm{\lambda}^\top h(g_{\mbf{u}^\star}(\mbf{i})),\\
m(\bm{\lambda}) &= \min_{\mbf{i}} \left( c(\mbf{i}) + \bm{\lambda}^\top h(g_{\mbf{u}^\star}(\mbf{i})) \right).
\end{align*}
By assumption, $\mathcal{L}$ is convex.  
Let $\mbf{i}'$ be a minimizer of $\mathcal{L}(\mbf{i}, \bm{\lambda}')$ for a given $\bm{\lambda}'$.  The core idea of the dual ascent method is that $h(g_{\mbf{u}^\star}(\mbf{i}'))$ is a (sub)gradient of $m$ at $\bm{\lambda}'$ because $\forall \bm{\lambda}$, we have:
\begin{align*}
m(\bm{\lambda}) &= \min_{\mbf{i}} \mathcal{L}(\mbf{i}, \bm{\lambda}),\\
&= \min_{\mbf{i}} c(\mbf{i}) + \bm{\lambda}^\top h(g_{\mbf{u}^\star}(\mbf{i})),\\
&\leq c(\mbf{i}') + \bm{\lambda}^\top h(g_{\mbf{u}^\star}(\mbf{i}')),\\
&= c(\mbf{i}') + \bm{\lambda}^{'\top} h(g_{\mbf{u}^\star}(\mbf{i}')) + (\bm{\lambda} - \bm{\lambda}')^\top h(g_{\mbf{u}^\star}(\mbf{i}')),\\
&= m(\bm{\lambda}') + (\bm{\lambda} - \bm{\lambda}')^\top h(g_{\mbf{u}^\star}(\mbf{i}')),
\end{align*}
where the last step follows because $\mbf{i}'$ is a minimizer of $\mathcal{L}(\mbf{i}, \bm{\lambda}')$.

Since $h(g_{\mbf{u}^\star}(\mbf{i}'))$ is a subgradient of $m$, a small move from $\bm{\lambda}$ in the direction of any subgradient at $\bm{\lambda}$ decreases the distance to any maximizer of $m$ (i.e., decreases the distance to an optimal dual solution).  To see this, we define $\mbf{s}^{(k)} \coloneqq h(g_{\mbf{u}^\star}(\mbf{i}^{(k)}))$, i.e., a subgradient of $m$, and let $\tilde{\bm{\lambda}}^{(k)} = \bm{\lambda}^{(k-1)} + \epsilon \mbf{s}^{(k)}$ denote the \textit{unprojected dual update}. Respectively, we will let $\bm{\lambda}^{(k)} = \left[ \tilde{\bm{\lambda}}^{(k)} \right]_{\mathbb{R}^n_+}$.  

Since the projection operator is non-expansive, for any $\tilde{\bm{\lambda}} \in \mathbb{R}^n$ and $\bm{\lambda}' \in \mathbb{R}^n_+$, we have that $\left\Vert \left[ \tilde{\bm{\lambda}} \right]_{\mathbb{R}^n_+} - \bm{\lambda}' \right\Vert_2^2 \leq \Vert \tilde{\bm{\lambda}} - \bm{\lambda}' \Vert_2^2$.   This gives us the following:
\begin{align*}
    \Vert \bm{\lambda}^{(k)} - \bm{\lambda}^\star \Vert_2^2 &\leq \Vert \tilde{\bm{\lambda}}^{(k)} - \bm{\lambda}^\star \Vert_2^2 = \Vert \bm{\lambda}^{(k-1)} - \bm{\lambda}^\star \Vert_2^2 + \epsilon^2 \Vert \mbf{s}^{(k)} \Vert_2^2 + 2 \epsilon \mbf{s}^{(k) \top} (\bm{\lambda}^{(k-1)} - \bm{\lambda}^\star)
\end{align*}
Note that $m(\bm{\lambda}^{(k-1)}) \leq m(\bm{\lambda}^\star)$ by definition of the dual problem.  We have the following condition:
\begin{align*}
\epsilon^2 \Vert \mbf{s}^{(k)} \Vert_2^2 &<  2 \epsilon s^{(k) \top} (\bm{\lambda}^\star - \bm{\lambda}^{(k-1)}),\\
\epsilon &<  \frac{2 s^{(k) \top} (\bm{\lambda}^\star - \bm{\lambda}^{(k-1)})}{\Vert \mbf{s}^{(k)} \Vert_2^2},\\
0 < \epsilon &<  \frac{2 \left( m(\bm{\lambda}^\star) - m(\bm{\lambda}^{(k-1)}) \right)}{\Vert \mbf{s}^{(k)} \Vert_2^2} < \frac{2 m(\bm{\lambda}^\star)}{\Vert \mbf{s}^{(k)} \Vert_2^2}.
\end{align*}
Note that $\bm{\lambda}^\star$ is the dual optimal point and thus $m(\bm{\lambda}^\star) = \mathcal{L}(\mbf{i}^\star, \bm{\lambda}^\star)$.  The condition on $\epsilon$ works for any initial $\bm{\lambda}^{(0)} \succ \mbf{0}$, by definition since $m(\bm{\lambda}^\star) \geq m(\bm{\lambda}) : \forall \bm{\lambda}$:
\begin{align*}
0 < \epsilon &< \frac{2 m(\bm{\lambda}^{(0)})}{\Vert \mbf{s}^{(k)} \Vert_2^2}.
\end{align*}
Furthermore, $\mbf{s}^{(k)}$ gives $h(g_{\mbf{u}^\star}(\mbf{i}^{(k)}))$.  Note that $h(g_{\mbf{u}^\star}(\mbf{i}^{(k)}))$ is upper bounded by $h(g_{\mbf{u}^\star}(\mbf{i}^{(0)}))$, where $\mbf{i}^{(0)}$ is the zero incentive.  This gives the following, which completes the proof:
\[
0 < \epsilon < \frac{2 m(\bm{\lambda}^{(0)})}{\Vert h(g_{\mbf{u}^\star}(\mbf{i}^{(0)})) \Vert_2^2}.
\]
\end{proof}

\subsection{Proof of \autoref{thm:FOIO-convergence}} \label{apx:FOIO-convergence}

In the following, we prove \autoref{thm:FOIO-convergence}, which states that under \sref{Assumptions}{asm:slater} and \ref{asm:boundedNorms}, \FOIO (\autoref{alg:foio}) converges to the unique minimizer of \eqref{eq:sci} if the step sizes are defined as
\begin{equation}
    \epsilon_k = \frac{\gamma_k}{\Vert \bm{\psi}^{(k)} \Vert_2}, 
\end{equation}
where $\gamma_k$ is a square summable but not summable positive quantity, and $\bm{\psi}^{(k)}$ collects the (sub)gradients of $\mathcal{L}$ with respect to $i$ and $\bm{\lambda}$.

\begin{proof}[Proof of \autoref{thm:FOIO-convergence}]
Recall that the Lagrangian is given by $\mathcal{L}(\mbf{i}, \bm{\lambda}) = c(\mbf{i}) + \bm{\lambda}^\top h(g_{\mbf{u}^\star}(\mbf{i}))$.  We start by defining a set-valued mapping $\bm{\psi}$ that collects the (sub)gradients of $\mathcal{L}$ with respect to $\mbf{i}$ and $\bm{\lambda}$, respectively:
\[
\bm{\psi}(\mbf{i},\bm{\lambda}) = \begin{bmatrix}
    \partial_{\mbf{i}} \mathcal{L}(\mbf{i}, \bm{\lambda})\\
    - \nabla_{\bm{\lambda}} \mathcal{L}(\mbf{i}, \bm{\lambda})
\end{bmatrix} = \begin{bmatrix}
   \mbf{1} - \nabla_{\mbf{i}} g_{\mbf{u}^\star}(\mbf{i})^\top \mbf{R}^\top \bm{\lambda}\\
   - h(g_{\mbf{u}^\star}(\mbf{i}))
\end{bmatrix}.
\]
Note that an optimality condition for \eqref{eq:sci} is given by $\mbf{0} \in \bm{\psi}(\mbf{i}^\star, \bm{\lambda}^\star)$.  

We will henceforth denote $\mbf{z}^{(k)} = (\mbf{i}^{(k)}, \bm{\lambda}^{(k)})$ as the $k^\text{th}$ iterate of the primal and dual variables in \FOIO.  $\mbf{z}^\star = (\mbf{i}^\star, \bm{\lambda}^\star)$ denotes any optimal pair of variables (i.e., a saddle point), where $\mathcal{L}(\mbf{i}^\star, \bm{\lambda}) \leq \mathcal{L}(\mbf{i}^\star, \bm{\lambda}^\star) \leq \mathcal{L}(\mbf{i}, \bm{\lambda}^\star)$, and we have $h(g_{\mbf{u}^\star}(\mbf{i}^\star)) \preccurlyeq \mbf{0}$, $\ \ \mbf{0} \in \partial_{\mbf{i}} \mathcal{L}(\mbf{i}^\star, \bm{\lambda}^\star)$.  Finally, we will let $p^\star$ denote the optimal objective value.

\noindent The primal-dual update in \autoref{alg:foio} can be written compactly as:
\[
\mbf{z}^{(k+1)} = \mbf{z}^{(k)} - \epsilon_k \bm{\psi}^{(k)},
\]
where $\bm{\psi}^{(k)}$ denotes any element of the set $\bm{\psi} ( \mbf{i}^{(k)}, \bm{\lambda}^{(k)} ) $, and $\epsilon_k > 0$ is the $k^\text{th}$ step size.
For the purposes of convergence, we use the following step size rule:
\begin{align*}
\epsilon_k = \frac{ \gamma_k }{\Vert \bm{\psi}^{(k)} \Vert_2 }, \ \text{ where $\gamma_k$ is chosen such that }  \gamma_k > 0, \ \sum_{k=1}^\infty \gamma_k = \infty, \ \sum_{k=1}^\infty \gamma_k^2 = S < \infty.
\end{align*}
We will henceforth assume that there is a scalar $A$ that satisfies
\[
 A \geq \Vert \mbf{z}^{(1)} \Vert_2 \ \ \text{ and } \ \ A \geq \Vert \mbf{z}^\star \Vert_2,
\]
i.e., $A$ upper bounds the 2-norm of the incentive and dual variables.  Note that by \sref{Assumption}{asm:threshold} this is intuitively true for the incentive, since $\mbf{t}$ is effectively the maximum incentive any solution should consider.  From this, a similar bound immediately follows for $\bm{\lambda}$ since the Lagrangian dual \eqref{eq:lagrang} enforces that the optimal $\bm{\lambda}^\star$ penalizes the objective ``enough'' (based on the violation of the constraint $h(\cdot) \preccurlyeq 0$) to drive the optimization towards offering an incentive, which is at maximum $\mbf{t}$.  On this finite set, recall that by \sref{Assumption}{asm:boundedNorms}, the norms of the (sub)gradients of $c(\cdot)$ and $h(g_{\mbf{u}^\star}(\mbf{i}))$ are bounded (finite).
Due to the update rule, we have the following identity:
\begin{align*}
    \Vert \mbf{z}^{(k+1)} - \mbf{z}^\star \Vert_2^2 &= \Vert \mbf{z}^{(k)} - \mbf{z}^\star \Vert_2^2 - 2 \epsilon_k \bm{\psi}^{(k) \top} (\mbf{z}^{(k)} - \mbf{z}^\star) + \epsilon_k^2 \Vert \bm{\psi}^{(k)} \Vert_2^2, \\
    &= \Vert \mbf{z}^{(k)} - \mbf{z}^\star \Vert_2^2 - 2 \gamma_k \frac{\bm{\psi}^{(k) \top}}{\Vert \bm{\psi}^{(k)} \Vert_2^2} (\mbf{z}^{(k)} - \mbf{z}^\star) + \gamma_k^2.
\end{align*}
By summing over $k$ and rearranging, we have:
\begin{align}
    \Vert \mbf{z}^{(k+1)} - \mbf{z}^\star \Vert_2^2 + 2 \sum_{i=0}^k \gamma_i \frac{\bm{\psi}^{(i) \top}}{\Vert \bm{\psi}^{(i)} \Vert_2^2} (\mbf{z}^{(i)} - \mbf{z}^\star) =\Vert \mbf{z}^{(1)} - \mbf{z}^\star \Vert_2^2 + \sum_{i=0}^k \gamma_i^2 \leq 4A^2 + S \label{eq:convFO}
\end{align}
To show that the sum on the left-hand side is non-negative, we have:
\begin{align*}
    &\bm{\psi}^{(k) \top} ( \mbf{z}^{(k)} - \mbf{z}^\star ) = \left( \mbf{1} - \nabla_{\mbf{i}} g_{\mbf{u}^\star}(\mbf{i}^{(k)})^\top \mbf{R}^\top \bm{\lambda} \right) (\mbf{i}^{(k)} - \mbf{i}^\star) - h(g_{\mbf{u}^\star}(\mbf{i}^{(k)}))^\top (\bm{\lambda}^{(k)} - \bm{\lambda}^\star)
\end{align*}
Note that $\mbf{1}$ is a subgradient of $c(\mbf{i})$ at $\mbf{i}^{(k)}$.  By definition, we have $\mbf{1}^\top (\mbf{i}^{(k)} - \mbf{i}^\star) \geq c(\mbf{i}^{(k)}) - p^\star$.  For the constraint, we have that $\left( - \nabla_{\mbf{i}} g_{\mbf{u}^\star}(\mbf{i}^{(k)})^\top \mbf{R}^\top \bm{\lambda} \right)^\top (\mbf{i}^{(k)} - \mbf{i}^\star ) \geq h(g_{\mbf{u}^\star}(\mbf{i}^{(k)})) - h(g_{\mbf{u}^\star}(\mbf{i}^\star)) = h(g_{\mbf{u}^\star}(\mbf{i}^{(k)})) $. 

\noindent Using these facts, we have:
\begin{align*}
    \bm{\psi}^{(k) \top} ( \mbf{z}^{(k)} - \mbf{z}^\star ) \geq c(\mbf{i}^{(k)}) - p^\star + \bm{\lambda}^{\star \top} h(g_{\mbf{u}^\star}(\mbf{i}^{(k)}) = \mathcal{L}(\mbf{i}^{(k)}, \bm{\lambda}^\star) - \mathcal{L}(\mbf{i}^\star, \bm{\lambda}^\star) \geq 0,
\end{align*}
where the last inequality follows because $\mbf{i}^\star$ minimizes $\mathcal{L}(\mbf{i}, \bm{\lambda}^\star)$ over $i$.
Since both terms on the left-hand side of \eqref{eq:convFO} are non-negative, the following bounds immediately follow:
\[
\Vert \mbf{z}^{(k+1)} - \mbf{z}^\star \Vert_2^2 \leq 4A^2 + S \quad \quad 2 \sum_{i=0}^k \gamma_i \frac{\bm{\psi}^{(i) \top}}{\Vert \bm{\psi}^{(i)} \Vert_2^2} (\mbf{z}^{(i)} - \mbf{z}^\star) \leq 4A^2 + S.
\]

Since we assumed that $\Vert \mbf{z}^\star \Vert_2$ is bounded, the first inequality implies $\mbf{z}^{(k)}$ cannot be too far from the origin.  (i.e., there exists a $D$ where $\Vert \mbf{z}^{(k)} \Vert_2 \leq D \ \forall k$; for example, $D = A + \sqrt{4A^2 +S}$).  

By \sref{Assumption}{asm:boundedNorms}, the norm of subgradients on this set of incentives $\Vert \mbf{i}^{(k)} \Vert_2 \leq D$ is bounded, so it follows that $\Vert \bm{\psi}^{(k)} \Vert_2$ is bounded.
Note that the sum of $\gamma_k$ diverges.  Thus, for the second sum to be bounded, it needs to be the case that
\[
\lim_{k\to \infty} \frac{\bm{\psi}^{(k) \top}}{\Vert \bm{\psi}^{(k)} \Vert_2^2} (\mbf{z}^{(k)} - \mbf{z}^\star) = 0.
\]

Since $\Vert \bm{\psi}^{(k)} \Vert_2$ is bounded, this implies that the numerator\\ $\bm{\psi}^{(k) \top} ( \mbf{z}^{(k)} - \mbf{z}^\star )$ must go to zero in the limit.  (i.e., we cannot rely on $\Vert \bm{\psi}^{(k)} \Vert_2$ going to $\infty$).  Recall the following inequality:
\begin{align*}
    0 \leq  \mathcal{L}(\mbf{i}^{(k)}, \bm{\lambda}^\star) - \mathcal{L}(\mbf{i}^\star, \bm{\lambda}^\star) \leq \bm{\psi}^{(k) \top} ( \mbf{z}^{(k)} - \mbf{z}^\star ).
\end{align*}
Since  $\mathcal{L}(\mbf{i}^{(k)}, \bm{\lambda}^\star) - \mathcal{L}(\mbf{i}^\star, \bm{\lambda}^\star) \geq 0$, this implies that\\ $\lim_{k \to \infty} \mathcal{L}(\mbf{i}^{(k)}, \bm{\lambda}^\star) = p^\star$.  Finally, we have:
\begin{align*}
    p^\star &= \lim_{k \to \infty} \mathcal{L}(\mbf{i}^{(k)}, \bm{\lambda}^\star)\\
    &= \lim_{k \to \infty} c(\mbf{i}^{(k)}) + \lim_{k \to \infty} \bm{\lambda}^{\star \top} h(g_{\mbf{u}^\star}(\mbf{i}^{(k)})) = \lim_{k \to \infty} c(\mbf{i}^{(k)}).
\end{align*}

This completes the proof, and the iterates of \FOIO converge to the optimal solution of \eqref{eq:sci} under \sref{Assumptions}{asm:slater} and \ref{asm:boundedNorms}.
\end{proof}

\subsection{Proof of \autoref{thm:FOIO-convergence-bad-T}} \label{apx:FOIO-convergence-bad-T}

In the following, we prove \autoref{thm:FOIO-convergence-bad-T}, which states that under \sref{Assumptions}{asm:slater}, \ref{asm:lipschitz-objective}, and \ref{asm:lipschitz-constraint}, \FOIO's iterates with coarse gradient estimates $\widehat{\nabla}_i g_{\mbf{u}^\star}(\mbf{i})$ satisfy
\begin{equation}
\lim_{k \to \infty} \sup \Vert \mbf{i}^{(k)} - \mbf{i}^\star \Vert_2 = O(\epsilon \Vert ( \bm{\delta} \oslash \mbf{t'} ) - ( \bm{\delta} \oslash \mbf{t} ) \Vert_2 ), 
\end{equation}
where $\mbf{i}^\star$ is the optimal solution to \eqref{eq:sci} when $g_{\mbf{u}^\star}(\mbf{i}) = \overline{g}_{\mbf{u}^\star}( \mbf{i})$.

\begin{proof}[Proof of \autoref{thm:FOIO-convergence-bad-T}]
Recall the incentive (primal) update in \FOIO (\autoref{alg:foio}) on line 4, given an estimate of the gradient $ \widehat{\nabla} \mathcal{L} (\mbf{i}^{(k)}, \bm{\lambda}^{(k)})$ that is constructed based on  $\widehat{\nabla}_i g_{\mbf{u}^\star}(\mbf{i})$:
\begin{align}
    \mbf{i}^{(k+1)} = \mbf{i}^{(k)} - \epsilon \widehat{\nabla} \mathcal{L} (\mbf{i}^{(k)}, \bm{\lambda}^{(k)})
\end{align}
We obtain bounds on the error $\Vert \mbf{i}^{(k)} - \mbf{i}^\star \Vert$, where $\mbf{i}^\star$ is the true (feasible) minimizer of $\mathcal{L}$.  
First, note that the following holds for any $\mbf{t'}$:
\begin{align*}
    \widehat{\nabla} \mathcal{L}(\mbf{i}, \bm{\lambda}) &= \mbf{1} - \widehat{\nabla}_i g_{\mbf{u}^\star}(\mbf{i})^\top \mbf{R}^\top \bm{\lambda},\\
    &= \mbf{1} - \nabla_{\mbf{i}} g_{\mbf{u}^\star}(\mbf{i})^\top \mbf{R}^\top \bm{\lambda} + \left(\nabla_{\mbf{i}} g_{\mbf{u}^\star}(\mbf{i}) - \widehat{\nabla}_i g_{\mbf{u}^\star}(\mbf{i}) \right)^\top \mbf{R}^\top \bm{\lambda},\\
    &= \nabla \mathcal{L}(\mbf{i}, \bm{\lambda}) + \left(\nabla_{\mbf{i}} g_{\mbf{u}^\star}(\mbf{i}) - \widehat{\nabla}_i g_{\mbf{u}^\star}(\mbf{i}) \right)^\top \mbf{R}^\top \bm{\lambda}.
\end{align*}
With respect to the true (linear) $g_{\mbf{u}^\star}(\mbf{i})$, we have the following:
\begin{align*}
    \widehat{\nabla} \mathcal{L}(\mbf{i}, \bm{\lambda}) &= \nabla \mathcal{L}(\mbf{i}, \bm{\lambda}) + \left[ \text{diag}( -\bm{\delta} \oslash \mbf{t} ) + \text{diag}( \bm{\delta} \oslash \mbf{t'} ) \right]^\top \mbf{R}^\top \bm{\lambda},\\
    &= \nabla \mathcal{L}(\mbf{i}, \bm{\lambda}) + \left[ \text{diag}( \left[ \bm{\delta} \odot (\mbf{t} \oslash \mbf{t'}) - \bm{\delta} \right] \oslash \mbf{t} \right]^\top \mbf{R}^\top \bm{\lambda},\\
    &= \nabla \mathcal{L}(\mbf{i}, \bm{\lambda}) + O \left( \Vert ( \bm{\delta} \oslash \mbf{t'} ) - ( \bm{\delta} \oslash \mbf{t} ) \Vert_2 \right).
\end{align*}

This gives that the \FOIO primal update on line 4 (with coarse gradient estimates) simplifies to:
\begin{align}
    \mbf{i}^{(k+1)} &= \mbf{i}^{(k)} - \epsilon \left[ \nabla \mathcal{L}(\mbf{i}, \bm{\lambda}) + O \left( \Vert ( \bm{\delta} \oslash \mbf{t'} ) - ( \bm{\delta} \oslash \mbf{t} ) \Vert_2 \right)  \right], \\
    &= \mbf{i}^{(k)} - \epsilon \nabla \mathcal{L}(\mbf{i}, \bm{\lambda}) + O \left( \epsilon \Vert ( \bm{\delta} \oslash \mbf{t'} ) - ( \bm{\delta} \oslash \mbf{t} ) \Vert_2 \right). \label{eq:approxUpdateFOinexact}
\end{align}

Given this update, techniques used to establish the convergence rate of gradient descent~\cite{BV:04} directly show that the asymptotic iterates of \eqref{eq:approxUpdateFOinexact} are ultimately bounded w.r.t. the optimal solution $\mbf{i}^\star$ on the true (linear) $g_{\mbf{u}^\star}(\mbf{i})$ -- in particular, if $\mathcal{L}$ is $L$-smooth (recall \sref{Assumptions}{asm:lipschitz-objective} and \ref{asm:lipschitz-constraint}), then we obtain the bound in \eqref{eq:FOinexactbound} irrespective of the initial condition, completing the proof.
\end{proof}

\subsection{Proof of \autoref{thm:ZOconvergence-nontimevary}} \label{apx:ZOconvergence-nontimevary}

In the following, we prove \autoref{thm:ZOconvergence-nontimevary}, which states that under \sref{Assumptions}{asm:slater}, \ref{asm:lipschitz-objective}, \ref{asm:lipschitz-constraint}, \ref{asm:measurement-error}, and \ref{asm:ident}, for a continuous time model, \ZOIO (\autoref{alg:zoio}) produces iterates that satisfy
\begin{equation}
\lim_{\tau \to \infty} \sup \Vert \mbf{i}(\tau) - \mbf{i}^\star \Vert = O(\epsilon + \sigma^2 + e_y).
\end{equation}

\begin{proof}[Proof of \autoref{thm:ZOconvergence-nontimevary}]
Consider the regularized Lagrangian $\mathcal{L}_{p,d}$, which is a $C^3$ strongly convex, strongly smooth function.  The corresponding zero-order method is given by the following update, which is itself given on line 5 (\ZOIO, \autoref{alg:zoio}).
\begin{align}
    \mbf{i}^{(k)} &= \mbf{i}^{(k-1)} - \epsilon \widehat{\nabla} \mathcal{L}_{p,d}^{(k-1)},\\
    &= (1- \epsilon p) \mbf{i}^{(k-1)} - \epsilon \widehat{\nabla} \mathcal{L}^{(k-1)},
\end{align}
Note that we cannot expect convergence of $\mbf{i}^{(k)}$ to a point as $k \to \infty$.  However, we may obtain bounds on the error $\Vert \mbf{i}^{(k)} - \mbf{i}^\star \Vert$, where $\mbf{i}^\star$ is the true (feasible) minimizer of $\mathcal{L}$.  
According to \cite[Lemma 1]{Chen:20}, the primal update in continuous time is given by
\begin{align}
\mbf{i'}(\tau) = \epsilon \beta(\tau, \mbf{i}(\tau)), \label{eq:ODE}
\end{align}
where for any $\tau$ and $\mbf{i}$,
\begin{align}
\beta(\tau, \mbf{i}) \coloneqq - \bm{\zeta}(\tau) \bm{\zeta}(\tau)^\top \nabla \mathcal{L}(\mbf{i}) + O(\sigma^2).
\end{align}
The function $\beta$ is Lipschitz continuous in $\mbf{i}$, uniformly in $\tau$.  We now discretize the continuous-time model by considering integer multiples of the period $P$.  At the $(K+1)$th stage of \ZOIO, we have thus far computed $\mbf{i}(KP)$, and with this initial condition, $\{ \mbf{i}(\tau) : KT \leq \tau \leq (K+1)T\}$ is defined as the solution to \eqref{eq:ODE}.  We then define $\mbf{i}((K+1)P)$ as:
\begin{align}
\mbf{i}((K+1)P) = \mbf{i}(KP) + \epsilon \int_{KP}^{(K+1)P} \beta(\tau, \mbf{i}(\tau)) d\tau. \label{eq:contTimeUpdate}
\end{align}
This recursion can be approximated by the following gradient descent update:
\begin{align}
    \mbf{i}((K+1)P) = \mbf{i}(KP) - \epsilon P (\nabla \mathcal{L}(i) + s_K(i)), \label{eq:approxUpdateZO}
\end{align}
where $s_K(i) = O(\epsilon + \sigma^2)$.

To justify this, observe that the integral in \eqref{eq:contTimeUpdate} can be approximated.  Under the Lipschitz conditions (\sref{Assumptions}{asm:lipschitz-objective} and \ref{asm:lipschitz-constraint}) and boundedness of $\bm{\zeta}$ (\sref{Assumption}{asm:ident}), there is a constant $b_0 < \infty$ such that for any $\tau \in [KP, (K+1)P]$,
\begin{align*}
\Vert \mbf{i}(\tau) - \mbf{i}(KP) \Vert &\leq b_0 P \epsilon,\\
\Vert \beta(\tau, \mbf{i}(\tau)) - \beta(\tau, \mbf{i}(KP)) \Vert &\leq b_0 P \epsilon.
\end{align*}
This implies that for $\tau$ in the range $[KP, (K+1)P]$, we have
\begin{align*}
    \beta(\tau, \mbf{i}(\tau)) &= \beta(\tau, \mbf{i}(KP)) + O(\epsilon) = -\bm{\zeta}(\tau) \bm{\zeta}(\tau)^\top \nabla \mathcal{L}(\mbf{i}(KP)) + O(\sigma^2) + O(\epsilon)
\end{align*}
Under \sref{Assumption}{asm:ident}, it follows that
\begin{align*}
    \int_{KP}^{(K+1)P} \beta(\tau, \mbf{i}(\tau)) d\tau &= -P \left[ \nabla \mathcal{L}(\mbf{i}(KP)) + O(\sigma^2 + \epsilon) \right],
\end{align*}
which completes the approximation in \eqref{eq:approxUpdateZO}.

Under this approximation, techniques used to establish the convergence rate and stability of gradient descent~\cite{Nesterov:13} can be used to show that the asymptotic iterates associated with \eqref{eq:ODE} are ultimately bounded.  In particular, if $\mathcal{L}_{p,d}$ is $L$-smooth and $\mu$-strongly convex (where note that this follows by definition of the regularized Lagrangian), we obtain the uniform bound in \eqref{eq:ZObound} irrespective of the initial condition, completing the proof.
\end{proof}

\end{document}